\documentclass[aps,prd,reprint,groupedaddress]{revtex4-2}
\input epsf.sty
\usepackage{graphics,amsmath,amssymb,color}

\def\lromn#1{\uppercase\expandafter{\romannumeral#1}}

\begin{document}


\title{
Parity violating magnetization  at   neutrino pair emission \\
using trivalent lanthanoid ions
}


\author{H. Hara}
\affiliation{Research Institute for Interdisciplinary Science,
Okayama University \\
Tsushima-naka 3-1-1 Kita-ku Okayama
700-8530 Japan}

\author{A. Yoshimi}
\affiliation{Research Institute for Interdisciplinary Science,
Okayama University \\
Tsushima-naka 3-1-1 Kita-ku Okayama
700-8530 Japan}

\author{M. Yoshimura}
\affiliation{Research Institute for Interdisciplinary Science,
Okayama University \\
Tsushima-naka 3-1-1 Kita-ku Okayama
700-8530 Japan}


\date{\today}

\begin{abstract}
A new detection method  using  magnetization  generated
at triggered radiative  emission of  neutrino pairs  (RENP),
$ |e \rangle  \rightarrow | g  \rangle + \gamma
+ \sum_{ij}\nu_i \bar{\nu}_j $  (atomic de-transition from state $|e \rangle $ to
state $|g \rangle$ emitting sum of neutrino pairs $\sum_{ij}\nu_i\bar{\nu}_j$
accompanied by a photon $\gamma$),
 is investigated in order to determine unknown neutrino properties; 
absolute neutrino masses of $\nu_i$ and Majorana/Dirac distinction.
Magnetization associated with RENP events has
parity violating component intrinsic to 
weak interaction enforced by crystal field effect in solids, 
and greatly helps background rejection of quantum electrodynamic 
(QED) origin even when these backgrounds are amplified.
In proposed experiments
we prepare a coherently excited body of trivalent lanthanoid ions, Er$^{3+}$
(a best candidate ion so far found), 
doped in a transparent dielectric crystal.
The magnetic moment $\mu \langle \vec{S}\cdot\vec{k} \rangle/k $
arising from generated  electron spin  $\vec{S}$ 
parallel to trigger  photon direction $\vec{k}/k$ is parity odd,
 and is absent in QED processes.
The generated magnetic field of order nano gauss  
 is stored in crystals long after pair emission event  till  spin relaxation time.
An improved calculation method of coherent rate and angular distribution of magnetization
is developed in order to incorporate finite size effect
of crystal target  beyond the infinite size limit in previous calculations.

\end{abstract}


\maketitle

\section{Introduction}

Neutrino oscillation experiments, despite their remarkable discovery
and measurement of finite neutrino mass and mixing in interactions,
cannot determine remaining important neutrino properties and physical parameters;
absolute neutrino masses and Majorana/Dirac distinction.
These are key ingredients to explain the matter anti-matter imbalance in
cosmology and construct the unified theory of particle physics.
We proposed more than ten years ago an idea to solve
these issues by using atoms/ions of typical eV energy release
much closer to expected neutrino masses
 instead of atomic nuclei of typical MeV energy release.
A mechanism of rate enhancement is crucial for weak processes,
and we initiated  experiments that verify the principle of enhanced event rates 
in weak QED  (Quantum Electro Dynamics) process \cite{renp overview}.
The  enhancement region extends to a whole body of excited target
consisting of many atoms/ions within which macro-coherence prevails.
The spatial region of macro-coherence for multiple particle
emission is much larger than the
coherence region  of single photon emission  in related Dicke's 
super-radiance scheme \cite{dicke 1}, \cite{dicke 2},
which is limited by photon wavelength $1/k= \lambda/2\pi$.
Our achieved rate enhancement reaches of order $10^{18}$ close to
expectation in QED two-photon emission
from excited  vibrational state of para-hydrogen \cite{psr exp 1}, \cite{psr exp 2},
\cite{psr exp 3}.

The process we have considered for neutrino study is either 
radiative neutrino pair $\nu\bar{\nu}$ emission (RENP)
$|e\rangle \rightarrow | g \rangle + \gamma + \nu\bar{\nu} $ stimulated by trigger laser, or
Raman stimulated radiative neutrino pair emission (RANP) \cite{ranp}
in atomic de-excitation between two states, $|e\rangle \rightarrow |g \rangle$.
One measures either dependence on signal (stimulated by trigger) photon energy
 or angular distribution  to extract neutrino properties.
A large number of target atoms/ions
close to those in solid environment is required for realistic RENP experiments.
We shall consider a special crystal
in the present work; trivalent lanthanoid ion Er$^{3+}$ of low concentration doped in 
host crystals, known to have narrow radiative widths.

A potentially serious problem against RENP detection is QED backgrounds.
Measurement of parity odd quantity, as first discussed in  \cite{pv ysu},
  in triggered neutrino pair emission is of great use to distinguish
the process involving weak interaction from purely
QED processes, even if they are macro-coherently amplified.
Random QED backgrounds that mimic parity violating effects are separated, since
 parity violating quantities in QED average to zero in statistically meaningful data.
An interesting quantity of this nature in RENP
is magnetization arising from magnetic moments parallel to the trigger photon direction 
generated at  neutrino pair emission:
the expectation value in the final state $| f\rangle $ in RENP,
$g \mu_{B} \langle f|   \vec{S} \cdot  \hat{ k}| f \rangle $,
where $\hat k$ is the unit vector along the signal photon momentum $\vec{k}$
and $\vec{S}$ is the electron spin operator to be multiplied by $g \mu_B$, 
the $g$-factor times Bohr magneton for magnetization.
This quantity is parity  odd and time reversal (T) even.
The generated magnetization due to these magnetic moments persists in crystals till spin relaxation time,
of order msec or even larger in appropriate trivalent lanthanoid ions
doped in host crystals.
Measurement of an accumulated bulk quantity such as magnetization gives more freedom
in experiments than direct event detection that rarely occurs \cite{ect}.

Parity odd quantity emerges from interference term of parity even and odd amplitudes
in the weak hamiltonian $H_W$ of neutrino pair emission.
Writing in terms of the neutrino mixing matrix elements, $U_{ei}\,, i=1,2,3\,,
\nu_e = \sum_i U_{ei} \nu_i $, the weak hamiltonian density consists of
two terms of different parities, 
\begin{eqnarray}
&& 
H_W =
\frac{G_F}{\sqrt{2}}\, \Sigma_{ij}\,
\bar{\nu}_i \gamma^{\alpha}  (1 - \gamma_5)\nu_j \,
\bar{e}\left( \gamma_{\alpha} c_{ij} - \gamma_{\alpha} \gamma_5 b_{ij}
\right) e
\nonumber \\ &&
\hspace{0.5cm}
\equiv
\frac{G_F}{\sqrt{2}}\left( - {\cal N}_b^{\alpha}
\bar{e} \gamma_{\alpha} \gamma_5 e  + {\cal N}_c^{\alpha} 
\bar{e}\gamma_{\alpha}  e
\right)
\,,
\label{weak cc}
\\ &&
 {\cal N}_b^{\alpha} = \Sigma_{ij}\, b_{ij} \,
\bar{\nu}_i \gamma^{\alpha}  (1 - \gamma_5)\nu_j 
\,, 
\nonumber \\ &&
{\cal N}_c^{\alpha} = \Sigma_{ij}\, c_{ij} \,
\bar{\nu}_i \gamma^{\alpha}  (1 - \gamma_5)\nu_j 
\,,
\\ &&
b_{ij} = U_{ei}^* U_{ej} - \frac{1}{2}\delta_{ij}
\,, 
\nonumber \\ &&
\hspace*{-0.3cm}
c_{ij} = U_{ei}^* U_{ej} - \frac{1}{2} ( 1- 4 \sin^2 \theta_w) \delta_{ij}
= b_{ij} + 2 \sin^2 \theta_w \delta_{ij}
\,,
\end{eqnarray}
where $\nu_i$ denotes a neutrino of definite mass $m_i$ and
$\theta_w$ is the weak mixing angle.
Electron axial vector current couples to neutrino-pair emission current 
$ {\cal N}_b^{\alpha} $,
while electron vector current couples to pair emission current $ {\cal N}_c^{\alpha} $.
Parameters $b_{ij}\,, c_{ij}$ are known except CP phases that appear in $U_{ei}$, 
one phase in the Dirac neutrino case and three phases  in the Majorana case.
Our strategy of neutrino mass spectroscopy is to assume 
unknown value $m_1$ of smallest neutrino mass and CP phases for fixing all other variables 
from neutrino oscillation measurements,
and calculate observable quantities for experimental comparison.
This way one does not have to commit to any particular model
beyond the standard theory, except finite neutrino masses and CP violation phases.
If CP conservation is assumed, all $b_{ij}\,, c_{ij}$'s are real and already known,
leaving $m_1$ as a single unknown parameter.
Decomposition of flavor states $\nu_e$ into neutrino mass eigenstates $\nu_i$  
is achieved in our proposed experiments by
precision of laser frequencies, usually less than $\mu$eV, and
not by measurable accuracy of  photon energy in detectors that may be larger than 1 meV.

The dominant interference term of parity violation arises from 
product of spatial part of axial vector current 
of electron, 
 $\bar{e} \vec{\gamma}\gamma_5 e$  which is the spin operator $\vec{S}$
in the non-relativistic limit of atomic electrons, 
and spatial part of vector current, the velocity operator 
$\bar{e} \vec{\gamma} e \sim \langle  \vec{v} \rangle = \langle \vec{p} \rangle/m_e$
(nearly equivalent to position operator $\vec{r}$ times energy difference
between two involved atomic states, often used in atomic physics).
Matrix elements of spin $\langle \vec{S} \rangle $  are usually of order unity,
while those of velocity operator $\langle \vec{v} \rangle $
are less than $10^{-3}$, hence interference terms are at least smaller by $10^{-3}$ 
than rate $\propto  \langle \vec{S} \rangle^2$ of a parity even quantity.
We shall quantify this ratio for lanthanoid ion we use as a target.
Magnetization caused by $(ij)$
($ ij $ is abbreviation for $\nu_i \bar{\nu}_j $) 
neutrino-pair emission is proportional to
neutrino mixing parameters,
$\Re (b_{ij} c_{ij})$ (assuming CP conservation for simplicity),
while parity conserving quantity of RENP event rate
is proportional to $\Re (b_{ij}^2)$.

A great advantage of 4f electrons in lanthanoid ions is that they are not 
sensitive to environmental effects of host crystals due to
filled 5s and 5p electrons in outer shells working as a shield.
This circumstance   gives both  a narrow transition width and a
large spin relaxation time, the latter being very important to magnetization measurement.
Lanthanoid ions of low concentration doped in host crystals exhibit
paramagnetic property at room temperature.
For an experimental purpose as explained below,  we shall use 
Kramers degenerate  ions
in which two states of different time reversal quantum numbers $\pm$
are energetically degenerate.

We develop
 in the present work an improved calculation method of macro-coherent rates and
magnetization by incorporating finite size effect of crystal target in integration   over neutrino momenta.
In the previous works we took the infinite volume limit giving
rise to the momentum conservation, or more properly the phase matching condition.
But it is difficult to justify this infinite size limit for experimentally used size of targets,
of order $\approx 1 $ cm.
This improved method gives quantitatively different results of
rate and magnetization, although the phase matching condition
may be still useful as a technical guide.
Angular distribution of magnetization, the most important physical quantity
to neutrino mass spectroscopy, is greatly influenced by
change of this calculation method.

The measurement method we propose here is different from those in previous proposals:
detection of accumulated  magnetization remaining in target medium  \cite{ect}
 rather than measuring direct individual events that rarely occur.
If the accumulated magnetization is above the sensitivity level of detectors, for instance, 
a high quality SQUID, then 
the method is found to be  sensitive to absolute neutrino mass determination
and to  Majorana/Dirac distinction.
Proposed experiments can be conducted irrespective of  the nature of neutrino masses,
Majorana or Dirac,
and  experiments themselves can determine  whether neutrinos are of either type
 by measuring the shape and the magnitude
of angular distribution of generated magnetization.
The principle of this distinction is due to
existence  of  interference term
intrinsic to identical fermions (particle = anti-particle)
in the Majorana  neutrino case \cite{my-07}. In the Dirac neutrino case
anti-neutrinos are distinguishable, hence there is no interference.

The present paper is organized as follows.
In the next section we first present amplitude of individual atomic process,
and in the next second section we spell out improved calculation method
of phase factors over entire target atoms
incorporating finite size effect of target.
In Section \lromn4 we explain trivalent Er ion used as a target candidate and
present results of magnetization calculation.
Effect of crystal field is quantified and used in calculation of atomic matrix elements.
A host crystal was chosen from the point of rich available optical data 
necessary for detailed computations.
In Summary and Outlook of Section \lromn5 we mention  points to be studied
for realistic experimental design.
In Appendix we explain four-level optical Bloch equation and
its solutions necessary to estimate populations in energy levels
and coherence parameters.

We use the natural unit of $\hbar = c = 1$ throughout the present work
 unless otherwise stated.

\begin{figure*}[htbp]
 \begin{center}
 \centerline{\includegraphics{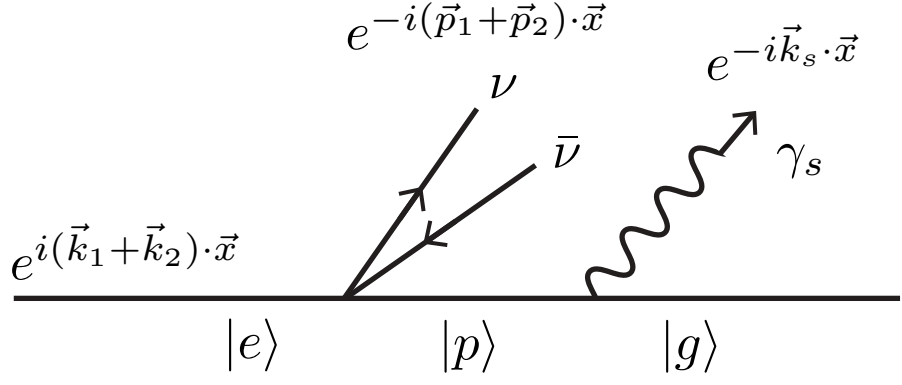}} \hspace*{\fill}
   \caption{
 Feynman diagram based on old-fashioned non-relativistic
perturbation theory of  RENP
$| e \rangle \rightarrow | g \rangle + \gamma_s  +  \nu\bar{\nu} $.
Relevant phase factors including  imprint to
state $|e\rangle $ given by the wave vector, $\vec{k}_1 + \vec{k}_2$,
at two-laser excitation are attached to ion and particle states.
}
   \label {feynman diagram}
 \end{center} 
\end{figure*}


\section{Probability amplitude of RENP process and parity violating magnetization}

A particular class of RENP process we consider in the present work
consist of individual events occurring each at a space point $\vec{x}$.
Its diagram is shown in Fig(\ref{feynman diagram}).
Individual atomic (or ionic) transition probability amplitude is given by
\cite{amplitude formula}
\begin{eqnarray}
&&
{\cal A}_{ij} e^{- i K_0 t + i \vec{K}\cdot \vec{x} }
\,, \hspace{0.5cm}
K_0 = \omega_1 +\omega_2 - \omega_s - E_1 - E_2
\,, 
\label {amplitude for each ion}
\\ &&
\vec{K} = \vec{k}_1 + \vec{k}_2-   \vec{k}_s - \vec{p}_1- \vec{p}_2
\,,
\\ &&
{\cal A}_{ij} =
- \frac{G_F}{\sqrt{2}} \frac{ (b_{ij} \vec{S}_{ep}+ c_{ij} \vec{v}_{ep}) \cdot \vec{{\cal N}}_{ij}
 (\vec{d}_{pg}\cdot \vec{E}_t + \vec{\mu}_{pg}\cdot \vec{B}_t)}
{\epsilon_{pg} - \omega_s - i \gamma_{ep}/2 }
\,, 
\nonumber \\ &&
\vec{{\cal N}}_{ij} =
\bar{\nu}_i \vec{\gamma}  (1 - \gamma_5)\nu_j 
\,.
\end{eqnarray}
Neutrino field operators $\nu_i\,, \bar{\nu}_j$ in  Introduction
should be replaced here by their plane wave functions, 
but we shall keep the same notation for simplicity.
We prepare excited state $| e\rangle $ by two-photon laser excitation
of their photon 4-momenta $( \omega_i, \vec{k}_i)\,, i = 1,2$
from the ground state $|g \rangle $.
Emitted signal photon $\gamma_s$ is stimulated by  trigger laser of photon
4-momentum $  ( \omega_t, \vec{k}_t)=( \omega_s, \vec{k}_s) $.
$(E_i = \sqrt{\vec{p}_i^2 + m_i^2},\vec{p}_i) \,, i = 1,2\,,$ are
 4-momenta of emitted neutrino pairs with their masses given by $m_i$.
The second order perturbation theory originally gives an energy denominator,
$1/(- \epsilon_{ep} +E_1 + E_2)$, which is transformed as above by using the exact rule of  energy conservation $\epsilon_{eg} (= \omega_1 + \omega_2) = \omega_s
+ E_1 + E_2$, in order to eliminate neutrino energy dependence and
make explicit dependence on the  signal photon energy $\omega_s$.
Two transition dipoles, magnetic ($\vec{\mu}_{pg}$) and electric
($\vec{d}_{pg}$), are included in trigger field coupling $\propto \vec{E}_t, \vec{B}_t$
where  laser intensity $I_t$ is given by
the power Watt divided by an area, being related to field strength
by $|\vec{E}_t |= | \vec{B}_t| = \sqrt{I_t} $.

\begin{figure*}[htbp]
 \begin{center}
 \epsfxsize=0.5\textwidth
 \centerline{\epsfbox{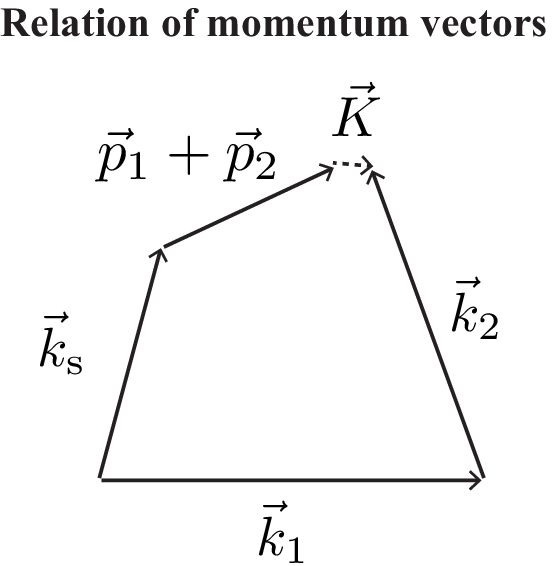}} \hspace*{\fill}\vspace*{1cm}
   \caption{ 
Relation of momentum vectors and definition of $\vec{K}$  with
fixed magnitudes, $ k_1= 1.286 \,, k_2=1.109 \,, k_s = 0.805$ in eV unit
for proposed Er$^{3+}$ experiment.
}
   \label {er quadrangle}
 \end{center} 
\end{figure*}

In Fig(\ref{er quadrangle}) we show momentum relation for target ion
we shall consider in the present work.
Three momenta, $\vec{k}_1\,, \vec{k}_2\,, \vec{k}_s=\vec{k}_t$, those for excitation and
trigger, and we assume 
by experimental design
 that they are in a plane, while individual neutrino momenta, 
$\vec{p}_i\,, i = 1,2$, may be out of this plane.

An implicit assumption  for RENP amplitude is to be noted:
emitted signal photons do not suffer from scattering or any reaction
with surrounding atoms in solid after their emission.
Even if there are interactions with surrounding atoms leaving modifications of amplitude
formula, the subsequent formula  for magnetization is related to RENP amplitude here,
since magnetization is generated immediately after RENP and prior to interactions.

When the trigger laser frequency 
$\omega_t = \omega_s$ is tuned to the resonance energy $\epsilon_{pg}$,
the amplitude becomes
\begin{eqnarray}
&&
{\cal A}_{ij} =
- i \sqrt{2} G_F h_r (b_{ij} \vec{S}_{ep}+ c_{ij} \vec{v}_{ep}) \cdot \vec{{\cal N}}_{ij}
\,, 
\nonumber \\ &&
h_r = 
\frac{ 
 \vec{d}_{pg}\cdot \vec{E}_t + \vec{\mu}_{pg}\cdot \vec{B}_t} {\gamma_{ep} }
\,.
\label {renp amplitude}
\end{eqnarray}
A remarkable feature of this formula  is that the RENP amplitude ${\cal A}_{ij}$
 is equal to neutrino-pair decay amplitude of $ |e \rangle \rightarrow | p\rangle
+ \nu_i \bar{\nu}_j$, simply multiplied by $2 h_r$ in which
effect of stimulated photon emission is coded.

Let us consider in more detail the factor $h_r $ multiplying the neutrino-pair emission amplitude,
which may be related to various observable quantities.
We consider unpolarized target and linear polarization of
trigger laser consisting of an equal mixture of two circular polarizations.
In this case
\begin{eqnarray}
&&
h_r =
\sqrt{ \frac{\pi}{\epsilon_{pg}^3 } } (\sqrt{ \gamma_{pg}^{{\rm MD} } }+ 
\sqrt{ \gamma_{pg}^{{\rm ED} }}) \frac{ \sqrt{I_t} }{\gamma_{ep} }
\,.
\end{eqnarray}
In the case of trivalent Er ion scheme later proposed, this combination of parameters is 
numerically estimated as
\begin{eqnarray}
&&
h_r \sim
0.748} \times 10^3 \sqrt{\frac{I_t }{10 {\rm mW cm}^{-2}}
\,,
\end{eqnarray}
if one takes radiative decay rate, 57.3 sec$^{-1}$, for $\gamma_{ep}$
\cite{e lifetime}.
If an inhomogeneously broadened width of order 
10 $\sim$ 100 MHz $\times 2\pi$ is taken for
$\gamma_{ep}$, this  number decreases to 
$0.69 \times (10^{-3} \sim 10^{-4})
\sqrt{I_t/10 {\rm mW cm}^{-2}}$.
We propose to use lowest Stark levels in each lanthanoid J-manifold, since
uncontrollable phonon relaxation is minimal for transitions between these levels.
We may leave this uncertainty by inserting $(57.3 {\rm sec}^{-1}/\gamma_{ep})^2$
in subsequent formulas of rate and magnetization.

Neutrino variables, their helicities and momenta, are extremely difficult
to measure, especially at these low energies,
hence one integrates over these unobservables in atomic experiments.
Neutrino helicity sum gives both  for electron spin-current and for velocity current,
\cite{renp overview}
\begin{eqnarray}
&&
\sum_{h_1, h_2} | \vec{{\cal N}}_{ij}\cdot \vec{j}_e |^2 = 
\frac{1}{2} (1 - \frac{\vec{p}_1\cdot\vec{p}_2}{E_1E_2} -\delta_M 
\frac{m_1 m_2}{ 4 E_1E_2} ) 
\vec{j}_e\cdot\vec{j}_e  
\nonumber \\ &&
+  
\frac{\vec{p}_1\cdot\vec{j}_e \vec{p}_2\cdot\vec{j}_e  }{E_1 E_2}
+ \cdots
\,,
\label {helicity summed sq amp}
\end{eqnarray}
where dots $\cdots$ give irrelevant T-odd terms.
If CP-odd phases of neutrino mixing matrix elements are taken into account, T-odd quantities such as $\vec{M}\cdot \vec{k}_s\times \vec{k}_1 $
(up-down asymmetry of magnetization $\vec{M}$ relative to plane 
spanned by two wave vectors $\vec{k}_s\,, \vec{k}_1$) 
become observable. We neglected  T-odd terms for simplicity in the present work,
assuming real mixing matrix.

Electron matrix elements of $(ij)$ neutrino pair emission are given by
$\vec{j}_e = \bar{e}(\vec{\gamma}\, c_{ij} - \vec{\gamma}\gamma_5 \, b_{ij})e $ 
with $e, \bar{e}$ being replaced by relevant  atomic wave function
and conjugate of bound electrons states, $|p\rangle\,, |e \rangle$.
$\delta_M = 0$ for Dirac neutrino and $\delta_M =1 $ for Majorana neutrino.
Directional average over electron current $\vec{j}_e$ may be taken
for unpolarized targets, which gives the last term  
in eq.(\ref{helicity summed sq amp}) of the form,
$\vec{p}_1\cdot\vec{p}_2/(3 E_1 E_2)$.
Neutrino variable dependence in $|{\cal A}_{ij}|^2$ is then
a function of $\vec{p}_1\cdot\vec{p}_2$
depending on the opening angle $\theta_{12}$ and magnitudes $p_i\,, i = 1,2$.

Presence of crystal field  is important
to make it possible to measure parity violating effect of fundamental interaction.
Both of RENP initial and final states, $|e \rangle$ and $ | f\rangle$, are eigenstates of QED interaction,
including static Coulomb interaction of target ion with host crystal ions, which is a crystal field effect.
Relevant parity violating quantity of our interest $\langle f | \hat{k}\cdot \vec{S}_e | f \rangle$
is transformed by parity operation $P$ to $- \langle f |P^{-1} \hat{k}\cdot \vec{S}_e P| f \rangle$.
Hence crystal field effect must produce parity mixture in the state $| f \rangle$;
$|f \rangle = |f\rangle_+ + | f\rangle_-$ such that 
$\langle f | \hat{k}\cdot \vec{S}_e | f \rangle = _+\!\langle f|\hat{k}\cdot \vec{S}_e | f \rangle_-
+  _-\!\langle f|\hat{k}\cdot \vec{S}_e | f \rangle_+$ in order to have non-vanishing
value of 
$\langle f | \hat{k}\cdot \vec{S}_e | f \rangle =
 - \langle f |P^{-1} \hat{k}\cdot \vec{S}_e P| f \rangle$.

The crystal field $V_c$ acting on a lanthanoid 4f$^n$ ion contains
mixture of parity different states due to interaction with surrounding host crystal ions:
by decomposing state vector into $| 4 f \rangle = | 4f\rangle_0 + \delta |4f \rangle$
with $ | 4f\rangle_0 $ defined as state without crystal field, one has
\begin{eqnarray}
&&
\delta |4f \rangle \simeq \left(\frac{\partial V_c}{\partial \vec{r}}\right)_{\vec{r} = 0}\cdot
\langle 5 d | \vec{r} | 4f \rangle\,  | 5d \rangle
\,,
\end{eqnarray}
as pointed out by \cite{van vleck}.
This peculiar situation, 
which does not occur for isolated atoms in the free space, 
arises naturally when lanthanoid ions are placed in crystals.
It is known that low-lying
lanthanoid ions of 4f$^n$ system are ordinarily activated by magnetic dipole transitions,
but can often simultaneously
have forced electric dipole transitions of different parity, being caused by
crystal field \cite{van vleck}: its calculation method was
 formulated in \cite{judd-ofelt 1}, \cite{judd-ofelt 2}.

\section{Improved calculation method of macro-coherent RENP rate}

We re-examine and improve
 the calculation method of macro-coherent RENP rate and persistent magnetization,
directly summing over  plane-wave factors  of relevant absorbed and emitted particles:
eq.(\ref{amplitude for each ion}) summed over ion positions $\vec{x}$.
Suppose  that one wants to  determine energy
and momentum to an accuracy level, $( 100 \sim 1) \mu $eV,
whose inverse corresponds to $6.6 \times (10^{-12} \sim 10^{-10})$ sec in time
and $ 2 \sim 200$ cm in length.
Time scale given here is much smaller than the usual measurement time
of atomic experiments,
which implies that the Fermi golden rule of the infinite time limit
should be valid and the energy conservation holds.
The length scale, on the other hand, is larger than a typical spatial region  
of actual experiments, $< 1$ cm.
Hence one cannot take the infinite size limit giving the momentum conservation
of macro-coherent rates.
What is achieved in the present improved method is
incorporation of finite size effect by which it
becomes possible to deal with intricate double resonance
region of dynamical variables.

The squared amplitude relevant for rate and observables is decomposed into a sum of parity-even and parity-odd terms. Differential and total rates arises from
 parity-even part, while magnetization discussed in the present work arises from parity-odd part.
In the present section we discuss parity-even rate and in later sections we
proceed to parity-odd contributions which shall be denoted by notations such
as $I_{PV}$.
We investigate total event rate emerging from $N = n V$ atoms or ions in
a target volume of cylinder, $V= \pi R^2 L$, irradiated by excitation and trigger lasers.
Event of an individual target atom/ion at a position $\vec{x}$ contributes to
the total amplitude a piece proportional to product of plane waves of
absorbed and emitted particles,
$e^{i \vec{K}\cdot \vec{x}} {\cal A}_{ij}$ ($\vec{K} =$ sum of 
an imprint wave vector + relevant momenta of  emitted particles.

In solids momentum vectors should be multiplied by refractive indexes of host crystals
denoted by $n_r$.),
with the atomic amplitude ${\cal A}_{ij} $ independent of $\vec{x}$ position.
We need to calculate an elementary integral of  squared amplitude over
neutrino momenta,
\begin{eqnarray}
&&
\int \frac{d^3 p_1 d^3p_2}{(2\pi)^6} 2\pi \delta (K_0)\, |{\cal A}|^2 \, I
\,, 
\nonumber \\ &&
I = (|\chi| \, n)^2\,
| \int_V d^3 x\, 
\exp[ i \vec{K}\cdot \vec{x}] |^2
\,,
\label {def space integral}
\end{eqnarray}
for the triggered process $| e\rangle  \rightarrow | g\rangle + \gamma_s + \nu\bar{\nu} $.
Here $n$ is the target ion number density assumed constant.
The averaged coherence $|\chi|$ may be estimated by solving
an optical Bloch equation, as discussed in Appendix.

In the previous simplified treatment \cite{renp overview}, \cite{ranp}, 
\cite{ect}, 
we took the infinite volume limit which gives the space integral of the form,
\begin{eqnarray}
&&
| \int_V d^3 x\, \exp[ i \vec{K}\cdot \vec{x}] |^2 
=  V\,(2\pi)^3 \delta ( \vec{K} )
\,.
\end{eqnarray}
How this result is changed for a large, but finite volume is
a subject of this section.

For a target region of cylinder, radius $R$, length $L$, and its volume $V= \pi R^2 L$
(having in mind $R\ll L = O(1)$ cm),
the spatial integration of phase factors in eq.(\ref{def space integral}) gives
\begin{eqnarray}
&&
\int_V d^3 x \exp[ i \vec{K} \cdot \vec{x}] = A_1 A_2
\,, \hspace{0.5cm}
A_1 = \frac{ 2\sin \frac{K_z L }{2} } { K_z} 
\,, 
\nonumber \\ &&
\hspace*{0.5cm}
A_2 = 2\pi \frac{R}{ K_{\perp} } J_1 (K_{\perp}R )
\,,
\\ &&
 | A_1 A_2|^2 =   V^2 U(\frac{ K_z L}{2}) W(K_{\perp}R) 
\,, \hspace{0.5cm}
U(x) = (\frac{\sin x}{x} )^2
\,, 
\nonumber \\ &&
\hspace*{0.5cm}
W(y) = (2 \frac{J_1(y)}{y} )^2
\,,
\end{eqnarray}
where $K_{\perp} (K_z)$ is the magnitude of  transverse (longitudinal)
 component  to cylinder axis of excitation (parallel to the first excitation laser photon $ \vec{k}_1$).
$J_1(z)$ ($\sim z/2$ as $z\rightarrow 0$) is the Bessel function of the first order.

Both functions $U(x)\,, W(y)  $  
decrease as arguments $|x|$ or $ |y|$ increases, showing damped oscillatory behaviors.
It is difficult to accurately incorporate these damped oscillations in numerical integration
over neutrino momenta, $\int d^3p_1 d^3p_2$.
We thus replace these by a smoothed out function, either 
Lorentzian function or Gaussian function.
It was found that Lorentzian functions are easier to handle from the point
of numerical computations.
Difficulty of numerical simulations based on Gaussian approximation arises
from its rapid decrease in large argument region.
Small, but non-negligible contributions in signal light angular distribution come from
a combination of one
peak region and the other tail region of Gaussian functions,
which greatly enhances computer tension.

The  neutrino-pair phase space integral necessary
for  rate calculation of RENP process,
$| e\rangle \rightarrow | g\rangle + \gamma_s + \nu\bar{\nu} $, 
depends on how spin matrix element $\vec{S}_{ep}$
(and smaller $\vec{v}_{ep}$, too) in the amplitude, 
eq.(\ref{renp amplitude}), is oriented.
In order to simplify unnecessary complications,
we shall assume for the sake of discussion here the case
in which contributions proportional to $\vec{S}_{ep} \cdot \vec{p}_1 \,
\vec{S}_{ep}\cdot \vec{p}_2$ averages out to give vanishing contribution.
More realistic spin and velocity orientations shall be discussed in Section 4.
Assuming the azimuthal symmetry, RENP rate is given by
\begin{eqnarray}
&&
\frac{ h_r^2 G_F^2}{(2\pi)^3} (nV)^2  |\chi|^2\, |\vec{S}_{ep}|^2
 \sum_{ij} {\cal F}_{ij}
\,, 
\\ &&
{\cal F}_{ij} = 
\int dE_1 \int dE_2 \,  \delta (E_1 + E_2 - \epsilon_{ep}) \,
\sqrt{E_1^2 - m_i^2} \sqrt{E_2^2 - m_j^2}  
\nonumber \\ &&
\cdot \int_{-\pi}^{\pi}  d\theta_1 
\int_{-\pi}^{\pi}  d\theta_2 \, | \sin\theta_1 \sin \theta_2 |
 \, U_0(\frac{K_{z} L}{2}) W_0(K_{\perp} R) \, {\cal R}_{ij}
\,, 
\label {rate formula 0}
\\ &&
U_0(x) = \frac{1}{1 + (x/x_l)^2 }
\,, \hspace{0.5cm}
x_l = 1
\,, \hspace{0.5cm}
 W_0(y) =  \frac{1}{1 + (y/y_t)^2 }
\,, 
\nonumber \\ &&
\hspace*{0.5cm}
y_t = \frac{32}{3\pi^2} \simeq 1.081
\,,
\\ &&
x =
\frac{L }{2} \left(p_1 \cos\theta_1 + p_2 \cos \theta_2 
- w_l \right)
\,, 
\\ &&
y = R \left( p_1 \sin \theta_1 + p_2 \sin \theta_2
- w_t \right)
\,, 
\\ &&
w_l = (\omega_1 + \omega_2 \cos \theta_e - \omega_s \cos \theta_s) n_r
\,,
\\ &&
w_t = (\omega_2 \sin \theta_e -  \omega_s \sin \theta_s) n_r
\,,
\end{eqnarray}
with $n_r$ the index of refraction of host crystal ($\sim 1.45 $ in host crystal
of Section \lromn4).
$ {\cal R}_{ij} $ is a function of $(E_i,\vec{p}_i)\,, i = 1,2$ up to bi-linear orders,
and shall be explicitly given below in eq.(\ref{r ij}) for each neutrino $(ij)$ pair.
We introduced here notations for light  directions:
$\theta_e$ refers to direction of the second excitation laser relative to the first excitation laser, while $\theta_s$ (equal to the trigger one $\theta_t$) does to signal photon direction relative to the first excitation laser.
Lorentzian parameters $x_l, y_t$ were determined by
requiring that Lorentzian functions coincide with exact plane-wave integrals of
$U(x) = (\sin x/x)^2\,, W(y) = (2 J_1(y)/y)^2 $
in values at $x=0\,, y=0$ and integrated values.
All angles are measured from z-axis of the first excitation laser.

\begin{figure*}[htbp]
 \begin{center}
 \centerline{\includegraphics{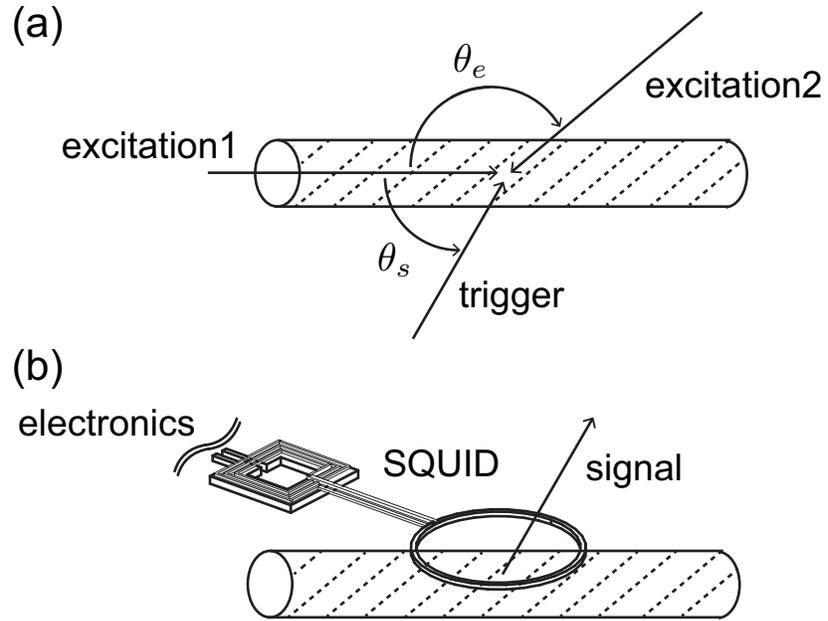}} \hspace*{\fill}
   \caption{ 
(a) Excitation scheme.
(b) Detection scheme  along with signal photon.
}
   \label {excitation and de-excitation}
 \end{center} 
\end{figure*}

\begin{figure*}[htbp]
 \begin{center}
 \centerline{\includegraphics{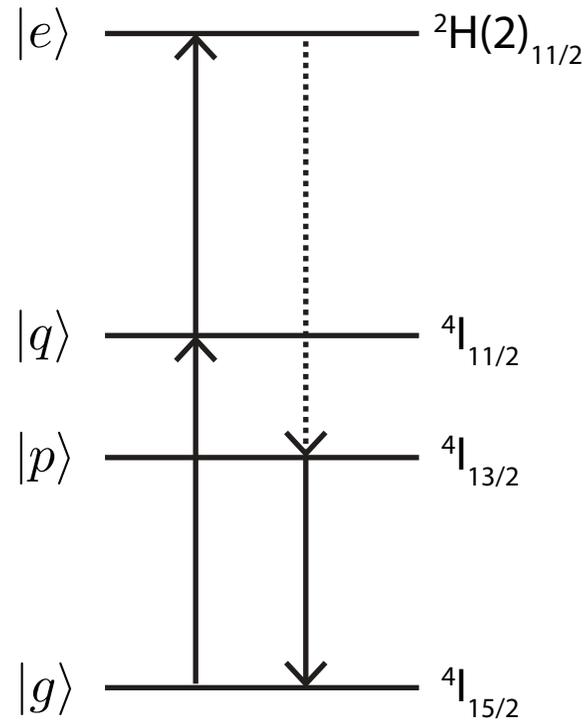}} \hspace*{\fill}
   \caption{
Excitation and trigger laser scheme for trivalent Er,
both depicted in arrowed solid lines,
and neutrino-pair emission depicted in arrowed dotted line.
}
   \label {excitation}
 \end{center} 
\end{figure*}

For large size values of $L\,, R$, two Loretzian functions, $U_0(x)\,, W_0(y) $, 
have sharp peaks in neutrino variables at
$x = 0\,, y=0 $, which restricts the region of largest contribution
in the neutrino phase space integral given by three independent variables,
$p_1, \theta_1,\theta_2$.
The largest region is given by $\vec{K}=0$
where $\vec{K}=  \vec{k}_1 + \vec{k}_2 - \vec{p}_1- \vec{p}_2 - \vec{k}_s $, and
momentum vector relation is depicted in Fig(\ref{er quadrangle}).
Thresholds of neutrino-pair production
are given by taking vanishing neutrino momenta $\vec{p}_i=0$.
The threshold condition gives  $w_t = 0\,, w_l=0$, or
\begin{eqnarray}
&&
\omega_2 \sin \theta_e -  \omega_s \sin \theta_s =0
\,, 
\nonumber \\ &&
\omega_2 \cos \theta_e - \omega_s \cos \theta_s = -  \omega_1
\,.
\label {aligned angles}
\end{eqnarray}
This determines both $\theta_s\,, \theta_e$:
in resonant trivalent Er scheme, energies $\omega_i\,, \omega_s$
are all fixed, and two angles are determined as
$\cos \theta_e = - 0.7838 \,, \cos \theta_s =  0.5182$ 
(142$^{\circ}$ and 59$^{\circ}$).
Thus, introduction of non-aligned second excitation laser of 
$ \theta_e \neq 0, \pi$ is required  to
experimentally approach neutrino-pair thresholds which are most sensitive
to Majorana/Dirac distinction.

Numerical pre-factor to be multiplied to 
$|\chi|^2\, |\vec{S}_{ep}|^2 \sum_{ij} {\cal F}_{ij}$  
is
\begin{eqnarray}
&&
C =
\frac{ 2 h_r^2 G_F^2}{(2\pi)^3} (nV)^2 
= 2.02 \times 10^{11}   \,{\rm sec}^{-1} {\rm eV}^{-5} 
\, \frac{ 57.3 {\rm sec}^{-1}}{\gamma_{ep} } \, 
\nonumber \\ &&
\times
\frac{ I_t}{ 10 {\rm mW cm}^{-2} }
(\frac{R }{0.5 {\rm cm} } )^4 (\frac{ L}{ {\rm cm}} )^2
(\frac{ n}{1.4 \times 10^{19} {\rm cm}^{-3} } )^2
\,,
\label {pre-factor}
\end{eqnarray}
with $V= 0.785 {\rm cm}^3 (R/0.5 {\rm cm})^2 L/{\rm cm}$.
The number in front is changed as
$2.0 \times 10^{11} \rightarrow 1.9 \times (10^5 \sim 10^4)$,
 if one takes 10 $\sim$ 100 MHz $\times 2\pi$ for $\gamma_{ep}$.

An important technical question concerns
how the infinite volume limit leading to
the momentum conservation is approached, ultimately giving a method of
how to determine the size  appropriate for realistic crystal growth.
This and related problems are postponed and discussed after we
solve how to compute magnetization in Section \lromn4.

\section{Trivalent Er ion scheme}

In the present work we shall restrict  to cases of neutrino pair emission,
in which two-photon
excitation from the ground state to excited state $|e\rangle $ is followed by
triggered RENP:
$| g \rangle + \gamma_1 +\gamma_2 \rightarrow | e \rangle;
| e \rangle  \rightarrow |g \rangle +\gamma_s + \nu\bar{\nu} $.
Experimental layout is shown in Fig(\ref{excitation and de-excitation}).
Excitation laser and trigger laser frequencies are matched to level spacings
of trivalent Er ion as shown in Fig(\ref{excitation}).

\begin{figure*}[htbp]
 \begin{center}
 \epsfxsize=0.6\textwidth
 \centerline{\epsfbox{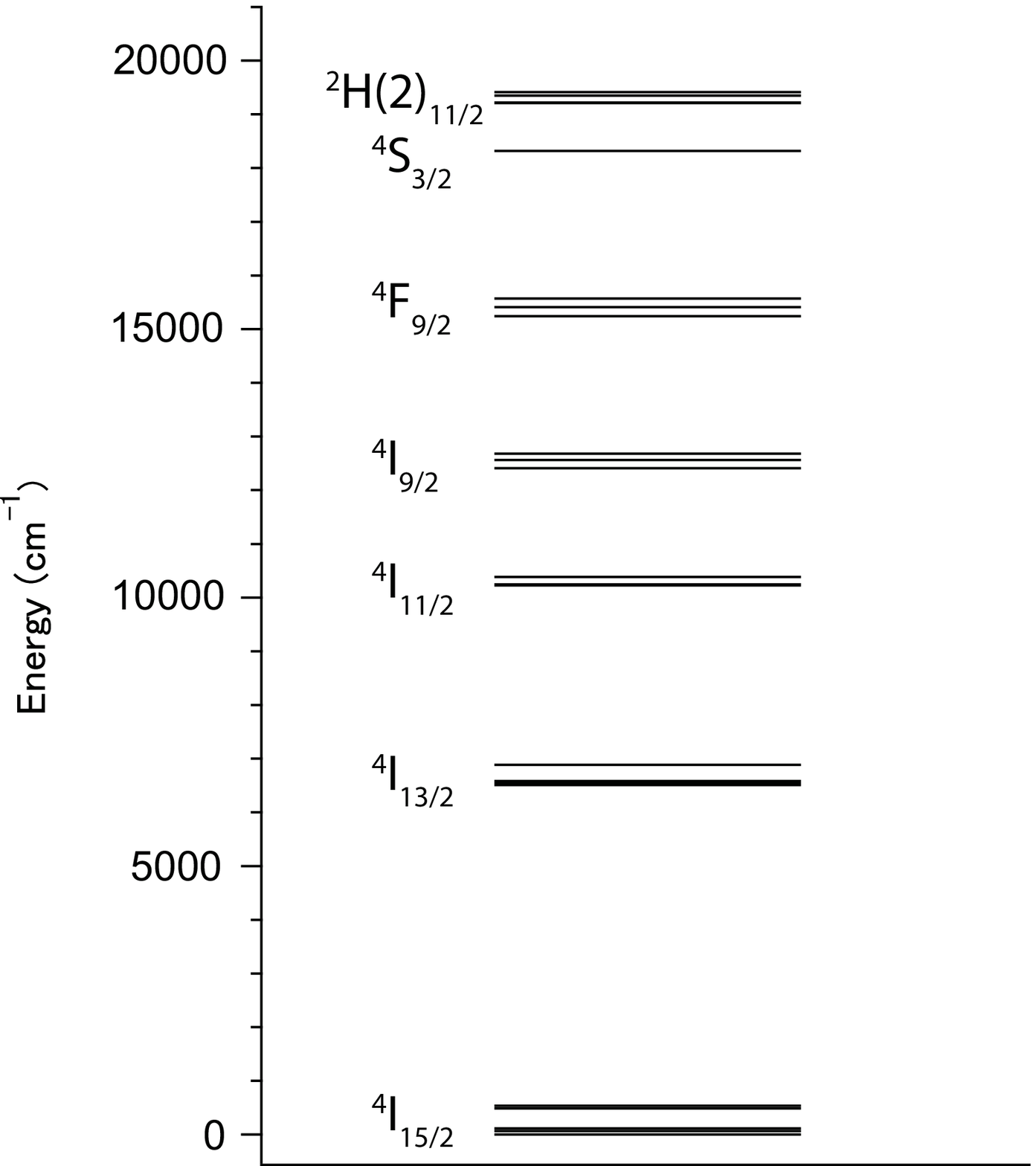}} \hspace*{\fill}\vspace*{1cm}
   \caption{ 
Low-lying trivalent Er levels doped in Cs$_2$NaYF$_6$ crystal.
Nearly identical, and unresolved with resolution of this figure, level structures appear
in YLF host crystal.
The energy conversion formula is $1\times 10^{4}$cm$^{-1} = 1.238\,$eV.
}
   \label {Er3+ levels}
 \end{center} 
\end{figure*}

Typical candidate targets we may consider are  trivalent Er and Ho ions which have
odd (11) and even (10) number of  4f valence electrons, respectively.
Trivalent 4f$^{11}$ Er ion has Kramers degeneracy, degeneracy due to
time reversal invariance, while Ho ion does not have this degeneracy
automatically.
Symmetry of host crystal may however induce accidental Kramers degeneracy
in non-singlets. This occurs in the ground state of Ho$^{3+}$
in a special host crystal of higher symmetry \cite{even 4f case}.

Lantanoid ions in the free space, or vacuum, are classified using the $LS$
coupling scheme, giving
  multiplets $^{2 S+1}\! L_J$ in which $J$ is the total angular momentum
combining the total spin $S$ and the total orbital $L$ angular momentum.
In the ground state of trivalent Er ion, $S= 3/2$ and $L= 6\,, J = 15/2$,
and is denoted by a term notation $^4$I$_{15/2}$.
These multiplets, called J-manifolds, are split into Stark states under crystal field
of host crystals, as illustrated  in Fig(\ref{Er3+ levels}).
Typical J-manifold splitting is of order 0.5 $\sim$ 1 eV,
while Stark splitting is of order several tens of meV \cite{lanthanoid spectroscopy}.
The number of separated Stark states in J-manifold is $J+1/2$ for  half-odd $J$,
each forming Kramers doublets, whose degeneracy may be
lifted, with amount of order $60 \mu$eV$B$/Tesla, by application of magnetic field.
Relaxation due to phonons in solids, which might become serious
obstacles against RENP,  is the smallest at the lowest
Stark state in each J-manifold, hence we shall use
the lowest Stark states of J-manifold in optical transitions of our scheme.

The choice of host crystals is important for success of RENP project.
We choose LiYF$_4$ (usually called YLF) crystal as host.
This crystal has a point group symmetry S$_4$ at the site of F,
and target ion Er$^{3+}$ replaces a fraction of F sites.
Energy levels of Stark states in doped crystal  Er$^{3+}$:YLF 
are well known, but we further need magnetic and electric dipole 
transition rates denoted by MD and ED.
For these we have to use available theoretical calculation of
different, but similar host crystal
Er$^{3+}$:Cs$_2$NaYF$_6$.
According to \cite{er3+}, calculated radiative decay rates of 10\% doped Er$^{3+}$ 
in host Cs$_2$NaYF$_6$ and other data are given in Table 1.

\begin{eqnarray*}
\hspace*{-0.5cm}
\begin{array}{ccccc}
{\rm initial }  & {\rm final }  & {\rm energy/eV } & {\rm  decay\; rate/sec}^{-1}  & v/S (10^{-6} )  \\ \hline
^4{\rm I}_{13/2} \rightarrow  &  ^4{\rm I}_{15/2} & 0.805 & 24.82^{{\rm MD}} + 2.46^{{\rm ED}} & 0.25 \\
^4{\rm I}_{11/2} \rightarrow & ^4{\rm I}_{15/2} & 1.286 & 4.13^{{\rm ED}} & \\
 & ^4{\rm I}_{13/2} & 0. 481 & 4.26^{{\rm MD}} + 0.42^{{\rm ED}}&0.15 \\
^2{\rm H}_{11/2} \rightarrow  &  ^4{\rm I}_{15/2} & 2.3946  &518.02^{{\rm ED}}  & \\
 & ^4{\rm I}_{13/2}  &  1.5894 &  49.43^{{\rm MD}}+7.91^{{\rm ED}}   & 1.2 \\
 & ^4{\rm I}_{11/2}  &  1.1091 &   5.58^{{\rm MD}} +4.73^{{\rm ED}} & 1.0\\
 &^4{\rm I}_{9/2} &  0.8318 &   0.49^{{\rm MD}} + 5.52^{{\rm ED}}& 2.7 \\
\hline
\end{array}
\end{eqnarray*}
\begin{center}
Table 1 Er$^{3+}$ levels and their radiative rates 
\end{center}

Coexistence of magnetic and electric dipole transitions, a necessary ingredient for magnetization measurement in RENP,
 is a manifestation of forced electric dipole moment  in 
lanthanoid crystals \cite{van vleck}, \cite{judd-ofelt 1}, \cite{judd-ofelt 2}.
The quantity $v/S$ in the last column contributes to ratio of magnetization to
rate which shall be precisely defined and calculated later.

Our RENP scheme of trivalent Er ion uses the following J-manifolds as in Fig(\ref{excitation}):
\begin{eqnarray}
&&
\hspace*{-0.5cm}
{\rm de-excitation}; 
\; | e\rangle =
^2\!{\rm H}_{11/2}( 2.3946 \,{\rm eV}) \rightarrow  
\nonumber \\ &&
|p \rangle = ^4\!{\rm I}_{13/2} ( 0.805\,{\rm eV} ) \rightarrow
|g \rangle = ^4\!{\rm I}_{15/2}(0\,{\rm eV})
\,, 
\label {er3+ scheme 1}
\\ &&
\hspace*{-0.5cm}
{\rm excitation}; \; ^4{\rm I}_{15/2}(0\,{\rm eV}) \rightarrow
\nonumber \\ &&
  | q\rangle = ^4\!{\rm I}_{11/2}(1.286\,{\rm eV}) 
\rightarrow ^2\!{\rm H}_{11/2}( 2.3946\,{\rm eV})
\,.
\label {er3+ scheme 2}
\end{eqnarray}
Hence excitation and trigger(=signal)  laser energies in eV unit are
\begin{eqnarray}
&&
\hspace*{-0.5cm}
\omega_1 = 1.286 \,, \hspace{0.5cm}
\omega_2 = 1.1086 \,, \hspace{0.5cm}
\omega_s = \omega_t = 0.805
\,.
\end{eqnarray}
The expected position of neutrino pair thresholds calculated from eq.(\ref{aligned angles})
 appears at signal (=trigger) direction given by
\begin{eqnarray}
&&
\cos \theta_s = \frac{ \omega_1^2 - \omega_2^2 + \omega_s^2}{ 2 \omega_1 \omega_s}
\sim 0.518 
\,,
\label {er3+ scheme 3}
\end{eqnarray}
corresponding to $\pm 1.026(\pm 59^{\circ}$) radians.
The second excitation laser should be irradiated with aligned angle at
\begin{eqnarray}
&&
\cos \theta_e = - \frac{ \omega_1^2 + \omega_2^2 - \omega_s^2}{ 2 \omega_1 \omega_2}
\sim -0.784
\,,
\label {er3+ scheme 4}
\end{eqnarray}
corresponding to $\pm 2.47(\pm 140^{\circ}$) radians.

Optical and neutrino-pair emission processes in solids are governed by the
selection rule of time reversal (T) quantum number:
its change must be T-odd, since two major single-photon operators,
electric dipole (E1) $ \vec{p}/m_e $ and magnetic dipole (M1) 
$\vec{\mu}$,  and neutrino-pair emission operators, $\vec{S} $
and $\vec{v}$,  are all T-odd.
(Use of the length gauge $e\vec{r}$ for E1 obscures this nature of T-oddness.)
On the other hand, two-photon excitation 
  is governed by T-even operator, hence
$|e \rangle $ and $|g\rangle $ are both T-even or both T-odd  \cite{t selection rule}.
The intermediate state $|p \rangle$ that participates in neutrino-pair emission
has different T quantum number from these. 

States constructed in the $LS$ coupling scheme
\begin{eqnarray}
&&
|\psi \rangle = a_J |J, M_J \rangle + b_{J'} | J', M_J'  \rangle + \cdots
\,,
\end{eqnarray}
are transformed under time reversal to
\begin{eqnarray}
&&
T |\psi \rangle = (-1)^{J+ M_J} a^*_J |J, - M_J \rangle 
\nonumber \\ &&
+ (-1)^{J'+ M_J'} b^*_{J'} | J', - M_J' \rangle + \cdots
\,.
\end{eqnarray}
In the leading order of our scheme, $(-1)^{J+  M_J} =1 $.
The second terms $\propto b_{J'}$ are sub-leading compared to
$\propto a_J$ terms.
States of $T = \pm $ are constructed as
\begin{eqnarray}
&&
\frac{1}{\sqrt{2}} (1 \pm T )|\psi \rangle \equiv |\psi^{\pm} \rangle
\,.
\end{eqnarray}
Matrix elements of T-odd operator $V$ are then given by
\begin{eqnarray}
&&
\langle \psi_2^- | V | \psi_1^+ \rangle = 
a_{J2}^* a_{J1} 
\langle J_2, M_{J2} | V | J_1, M_{J1}  \rangle
\nonumber \\ &&
\hspace*{0.5cm}
+ (-1)^{J_1 + M_{J1}}  a_{J2}^* a^*_{J1} \langle J_2,  M_{J2}   | V | J_1, - M_{J1}   \rangle
\,,
\end{eqnarray}
to leading orders.
The first term dominates over the second term due to smaller $\Delta M_J$.

Crystal field acting on trivalent Er ion 
(sum of Coulomb interaction caused by surrounding host ions)
splits J-manifold states into $J+1/2$ degenerate Kramers doublets,
consisting of T-even and -odd linear combinations of 
$\sum_{M_J} c_{M_J>0} (|J,  M_J \rangle \pm  (-1)^{J+M_J}  (|J, - M_J \rangle)$ 
with quantization axis
taken parallel to c-axis of tetragonal host crystal.
Magnetic quantum numbers of lowest Stark levels for relevant RENP  J-manifolds are
mixtures of different $M_J$ values, and
their weights are calculated using tabulated crystal field parameters 
\cite{judd-ofelt parameters}. 
Our calculation shows that dominant components have probabilities,
\begin{eqnarray}
&&
|e \rangle \; ^2{\rm H}_{11/2} (\Gamma_{78}\, {\rm of}\; S_4) 
\nonumber \\ &&
\; ; M_J = \pm \frac{1}{2}:  
0.1547
\,, \hspace{0.3cm}
M_J = \pm \frac{7}{2}:  0.2401
\,,
\\ &&
|p \rangle \; ^4{\rm I}_{13/2} (\Gamma_{56})
\nonumber \\ &&
\; ;M_J = \pm \frac{3}{2} :
0.2559 ;
\,, \hspace{0.3cm}
 M_J = \pm \frac{5}{2} : 0.1227
\,,
\\ &&
| g \rangle \; ^4{\rm I}_{15/2} (\Gamma_{78})\; ;M_J = \pm \frac{5}{2}: 0.2002
\,, \hspace{0.3cm}
M_J = \pm \frac{3}{2}:  0.1374
\,.
\nonumber \\ &&
\end{eqnarray}
$\Gamma_{78}\,, \Gamma_{56}$ denote irreducible representations of relevant
S$_4$ point group 
\cite{point group} of host crystal.
Neutrino pair emission at $|e \rangle \rightarrow | p\rangle$
followed by stimulated photon emission at $|p \rangle \rightarrow | g\rangle$
can occur, going through states of magnetic quantum number changes,
$M_J = \pm 1/2 \rightarrow \pm 3/2 \rightarrow \pm 5/2 $ by minimum steps of
$\Delta M_J =  \pm 1$.
This path gives the largest contribution.

We need to calculate matrix elements of
$\vec{S}_{ep}\,, \vec{v}_{ep}\,, \vec{\mu}_{pg} = 
\mu_B(\vec{J}+\vec{S} )_{pg} \,, \vec{d}_{pg} = i e \vec{v}_{pg}/\epsilon_{pg} $, 
 including necessary crystal field effect.
Calculation of spin matrix elements uses mixing amplitude times 3j-symbol.
For neutrino pair emission at $|e \rangle \rightarrow | p \rangle$,
\begin{eqnarray*}
&&
\langle p; \frac{13}{2}, \Gamma_{56} | S_+ | e; \frac{11}{2},  \Gamma_{78} \rangle
= 0.199 (= \sqrt{ 0.1547 \times 0.2559 })\,
\nonumber \\ &&
\cdot
 (-1)^{13/2- 3/2 } 
\left(
\begin{array}{ccc}
\frac{13}{2} & 1  &  \frac{11}{2} \\
-  \frac{3}{2} &  1  &  \frac{1}{2}
\end{array}
\right)
\langle p; \frac{13}{2} || S|| e; \frac{11}{2} \rangle
\,.
\end{eqnarray*}
For this choice of $M_J =1/2\,, M_J'=3/2$, other spin components, 
matrix elements of $S_0\,, S_-$ vanish.
This means that vector matrix elements are parallel to
crystal c-axis.

For estimate of $\vec{v}_{ep} = \pm i \vec{d}_{ep} \epsilon_{ep}/e$
we need to consider the crystal field effect of host crystals,
since parity forbids dipole elements between the same parity J-states.
The leading effect of host crystal surrounding target ion
is to introduce Coulomb-induced potential derivative of the form,
\begin{eqnarray}
&&
\left(\sum_l \vec{\nabla} V_C(\vec{r}_l - \vec{r}) \right)_{\vec{r} =0} \cdot 
\langle f | \vec{r} |i \rangle
\,, 
\\ &&
\left( \sum_l \vec{\nabla} V_C(\vec{r}_l - \vec{r}) \right)_{\vec{r} =0} 
= O( \frac{10{\rm eV}}{a})
\,,
\end{eqnarray}
with $a$ the lattice constant.
Thus, a direct product of two vector operators
for instance,
\begin{eqnarray}
&&
\frac{\left( \sum_l \vec{\nabla} V_C(\vec{r}_l - \vec{r}) \right)_{\vec{r} =0}  } { \Delta E}\,
\langle p; \frac{13}{2}, M' | ( r_0  v_+ + r_+ v_0)| e; \frac{11}{2},  M \rangle
\,,
\nonumber \\ &&
\end{eqnarray}
is non-vanishing,
where $\Delta E$ is a typical energy difference of 5d-4f ionic levels
(roughly of order 10 eV).
We defined components of vector operator $\vec{V}$ 
parallel to quantization axis as $V_0$, and
perpendicular components as $V_{\pm}$, depending on their angular momentum.
In Judd-Ofelt theory higher order terms $O(r^{2n+1})$ in crystal field expansion
and higher levels than 5d are effectively taken into account.

RENP amplitude is finally given in terms of reduced matrix elements
which are related to observables.
The following general formula for  calculation of reduced matrix element of
a vector operator is useful:
\begin{eqnarray}
&&
\hspace*{-0.3cm}
\sum_{m' q} | \langle j+1, m' | A_q|j, m\rangle |^2
= \frac{1}{2j+1} | \langle j+1 || A|| j \rangle |^2
\,.
\end{eqnarray}
For application to the spin operator, we first note
a relation in $LS$ coupling scheme, 
\begin{eqnarray}
&&
\hspace*{-0.5cm}
\langle J +1 |\vec{S}|J \rangle = \langle J +1 |\vec{S} + \vec{J} |J \rangle
= \langle J +1 |\vec{L} + 2\vec{S}|J \rangle
\,,
\end{eqnarray}
which gives, when multiplied by the Bohr magneton $\mu_B$,
magnetic dipole operator $\vec{M}$.
Hence,
\begin{eqnarray}
&&
| \langle p; \frac{13}{2} || S || e; \frac{11}{2} \rangle |^2  = 
\frac{ 12 }{\mu_B^2}
\sum_{m' q} | \langle p; \frac{13}{2}, m' | M_q|e; \frac{11}{2} , m\rangle |^2
\nonumber \\ &&
\hspace*{0.5cm}
= 12 \frac{ 3\pi \, \gamma_{ep}^{MD}} {\mu_B^2 \, \epsilon_{ep}^3}
\,.
\end{eqnarray}
Application of this formula to velocity operator requires crystal field effect, giving
\begin{eqnarray}
&&
| \langle p; \frac{13}{2} ||V_{{\rm crys}} v|| e; \frac{11}{2} \rangle |^2  = \frac{ 12 \epsilon_{ep}^2}{e^2}
\sum_{m' q} | \langle p; \frac{13}{2}, m' | d_q|e; \frac{11}{2} , m\rangle |^2
\nonumber \\ &&
\hspace*{0.5cm}
= 12 \frac{ 3\, \gamma_{ep}^{ED}} {4\, \alpha\, \epsilon_{ep}}
\,.
\end{eqnarray}
Numerically, we find for trivalent Er scheme,
\begin{eqnarray}
&&
\langle p; \frac{13}{2} || S || e; \frac{11}{2} \rangle \approx 3.47 
\,,\hspace{0.3cm}
\langle p; \frac{13}{2} || v|| e; \frac{11}{2} \rangle \approx 2.79 
\times 10^{-6}
\,.
\nonumber \\ &&
\end{eqnarray}
These give $\vec{S}_{ep}^2 \approx 0.0157$ for lowest Stark states  in our scheme.

In order to work out parity violating interference term,
we start from squared amplitude
when neutrino helicities are specified (not summed over), 
although neutrino helicities cannot be measured.
This calculation gives more insight than providing helicity summed squared amplitude
from the outset.
Neutrino pair emission of two plane-wave modes
specified by their momenta and helicities, $(\vec{p}_1 h_1)\,, (\vec{p}_2 h_2)$,
1 and 2 referring to anti-neutrino (distinguishable from neutrino only for Dirac case)
 and neutrino variables, respectively, 
gives   interference term of squared amplitude at $|e\rangle \rightarrow | p\rangle$
\cite{corrected my prd}
\begin{eqnarray}
&&
(I_{PV})_{12} = 2 \sum_{ij} \,\Re(v_i\, {\cal N}_{12}^i ({\cal N}_{12}^{j})^{\dagger}\,S_j ) 
=  (N_{12} -  \delta_M\, \frac{m_1 m_2}{ 16 E_1 E_2} )
\nonumber \\ &&
\cdot
\left\{ C_1 \vec{v}\cdot \vec{S}
+ C_2 \left( \hat{p}_1\cdot \vec{v}\, \hat{p}_2\cdot \vec{S}
 +  \hat{p}_1\cdot \vec{S} \, \hat{p}_2\cdot \vec{v}
\right) \right\}
\,,
\label {vs interference}
\\ &&
C_1 = (1 + h_1 h_2 \frac{ \vec{p}_1 \cdot \vec{p}_2}{p_1 p_2 } )
\,, \hspace{0.5cm}
C_2 = -  h_1 h_2 
\,, 
\nonumber \\ &&
N_{12} = \frac{1}{4}(1 + h_1 \frac{p_1}{E_1})(1 - h_2 \frac{p_2}{E_2}) 
\,, 
\\ &&
\vec{p}_1 \cdot \vec{p}_2 = p_1 p_2 \cos \theta_{12}
\,, \hspace{0.5cm}
\hat{p}_i = \frac{\vec{p}_i }{ p_i} 
\,,
\end{eqnarray}
where $\theta_{12}$ is the opening angle of emitted neutrino-pair,
and $ \hat{p}_i $ is neutrino propagation vector of unit length.

Two important limiting cases of quantity $(I_{PV})_{12}$ are listed in the following tables;
the case of relativistic (R) neutrinos and the case of non-relativistic (NR)
neutrinos.
The relativistic case gives dominant helicity combination of left-handed neutrino and
right-handed anti-neutrino (opposite helicity combination for
Majorana neutrino pair),
all other combinations giving vanishing contribution.
The non-relativistic case gives a different picture of nearly all non-vanishing
helicity combination, as shown in the table.

\vspace{0.5cm}
\hspace*{-0.3cm}
\begin{tabular}{|c|c|c|c|} \hline 
$(h_1, h_2) $ & $ \overline{C}_1 $ &  $\overline{C}_2 $ & comments \\ \hline \hline
$(1, -1) $ & $ 1-\cos\theta_{12} $ & $ 1 $ & common to Majorana and Dirac \\ \hline
$(1,1) $ & $0 $ & $0 $ &   common to MD\\ \hline
$(-1,1) $ & $ 0$ & $ 0$ &   common to MD\\ \hline
$(-1,-1) $ & $0 $ & $0 $ &   common to MD\\ \hline
Sum & $ 1-\cos\theta_{12} $ & $1$ &  common to MD \\ \hline
\end{tabular}
\begin{center}
Table 2 Contribution to $(I_{PV})_{12} $ of neutrino helicity combinations: relativistic case.
The following definition was used;
$\overline{C}_i = C_i  (N_{12} -  \delta_M\, \frac{m_1 m_2}{ 16 E_1 E_2} )\,, i = 1,2$.
\end{center}

\vspace{0.5cm}
\hspace*{-0.3cm}
\begin{tabular}{|c|c|c|} \hline 
$(h_1, h_2) $ & $ \overline{C}_1$ &  $ \overline{C}_2 $ \\ \hline \hline
$(1, -1) $ & $ \frac{1}{4}(1 +\frac{p_1 + p_2}{m}- \frac{\delta_M}{4} ) (1-\cos\theta_{12}) $ 
& $ \frac{1}{4} ( 1 + \frac{p_1+ p_2}{m} - \frac{\delta_M}{4} )$ 
  \\ \hline
$(1,1) $ & $\frac{1}{4}( 1 +\frac{p_1 - p_2}{m} - \frac{\delta_M}{4})
(1+\cos\theta_{12}) $ 
& $- \frac{1}{4} ( 1 + \frac{p_1- p_2}{m} - \frac{\delta_M}{4}) $ 
 \\ \hline
$(-1,1) $ & $\frac{1}{4}( 1 - \frac{p_1 + p_2}{m} - \frac{\delta_M}{4})
(1-\cos\theta_{12})  $ 
& $ \frac{1}{4} ( 1 - \frac{p_1+ p_2}{m} - \frac{\delta_M}{4}) $ 
  \\ \hline
$(-1,-1) $ & $ \frac{1}{4}(1- \frac{p_1 - p_2}{m} - \frac{\delta_M}{4})
(1+\cos\theta_{12}) $ 
& $ -\frac{1}{4} ( 1 - \frac{p_1- p_2}{m} - \frac{\delta_M}{4}) $ 
  \\ \hline
Sum & Dirac$ \frac{3}{4}({\rm Majorana}\; 1) $ & $ 0(0)$ 
\\ \hline
\end{tabular}
\vspace{0.5cm}
\begin{center}
Table 3 Contribution to $(I_{PV})_{12} $ of neutrino helicity combinations: 
non-relativistic case.
The factor $\delta_M = 0$ for Dirac neutrino and $=1$ for Majorana neutrino,
assuming an equal mass pair of $m_i = m\,, i=1,2$.
\end{center}

The helicity-summed interference term relevant to
actual experiments  is given by
\begin{eqnarray}
&&
\sum_{h_1, h_2} C_1 (N_{12} -  \delta_M\, \frac{m_1 m_2}{ 16 E_1 E_2} )
= 1 - \frac{ \vec{p}_1 \cdot \vec{p}_2}{ E_1 E_2 }
 - \delta_M\, \frac{m_1 m_2}{ 4 E_1 E_2}
\,, 
\nonumber \\ &&
\\ &&
\sum_{h_1, h_2} C_2 (N_{12} -  \delta_M\, \frac{m_1 m_2}{ 16 E_1 E_2} )
=   \frac{  p_1 p_2 }{  E_1 E_2} 
\,.
\label {h-summed pair current squared}
\end{eqnarray}
This result is equivalent to eq.(\ref{helicity summed sq amp}) in the preceding section.
Neutrino pair phase space integration over momenta
of $(I_{PV})_{12}$ further gives PV interference contribution,
\begin{eqnarray}
&& 
\hspace*{-0.5cm}
\tilde{F}_{PV} = \tilde{J}_1 \vec{v}\cdot \vec{S} + (v_i S_j + S_i v_j) \tilde{J}_2^{ij}
\,, \hspace{0.5cm}
\tilde{J}_1 = 
\\ &&
\hspace*{-0.5cm}
\int dE_1 \int dE_2 \delta (E_1 + E_2 - \epsilon_{ep} )
\int_{-\pi}^{\pi} d\theta_1 |\sin \theta_1|
\int_{-\pi}^{\pi} d\theta_2 |\sin \theta_2| 
\nonumber \\ &&
\cdot 
p_1 p_2  \, 
U_0(x) W_0(y) \cdot \left( E_1 E_2 - \vec{p}_1\cdot \vec{p}_2
- \delta_M m_1m_2
\right)
\,,
\label {tilde j1}
\\ &&
\tilde{J}_2^{ij} =  
\int dE_1 \int dE_2 \delta (E_1 + E_2 - \epsilon_{ep})
\int_{-\pi}^{\pi} d\theta_1 |\sin \theta_1| 
\nonumber \\ &&
\cdot
\int_{-\pi}^{\pi} d\theta_2 |\sin \theta_2| \,
p_1 p_2 \, U_0(x) W_0(y)  \, p_1^i p_2^j
\label {tilde j2}
\,.
\end{eqnarray}
This gives contribution from a single neutrino-pair, and
one should further add all six pair contributions with appropriate weights.

On the other hand,
RENP rate is given by parity conserving quantity taking dominant spin current contribution,
\begin{eqnarray}
&& 
\tilde{F}_{PC} \simeq \tilde{J}_1\vec{S}^2 + 2S_i S_j  \tilde{J}_2^{ij}
\,.
\end{eqnarray} 

Details of RENP rate and magnetization depend on how
crystals are placed relative to directions of excitation and trigger laser irradiation.
Crystal c-axis defines quantization axis and  matrix element
$\vec{S}_{ep}$ is oriented along this axis due to the selection rule,
$\langle J', M+1 | S_i | J, M \rangle =0 $ unless $S_i = S_+$.
For simplicity we take direction of the first laser excitation $\vec{k}_1$,
either parallel or orthogonal to c-axis.
There are two important cases of interest:
(1) c-axis is parallel to  $\vec{k}_1$,
(2) c-axis is not parallel to  $\vec{k}_1$, but in the plane spanned by two vectors, $\vec{k}_1\,, \vec{k}_2$,
(3) c-axis orthogonal to the plane  spanned by these two vectors.
We assume the signal wave-vector $\vec{k}_s$ in $\vec{k}_1\,, \vec{k}_2$ plane.
In the neutrino phase space integral of $\tilde{J}_2^{ij}$ for case (3) 
$\vec{p}_i \cdot \vec{S}_{ep}$ and $\vec{p}_i \cdot \vec{v}_{ep}$ give odd functions to integrands,
leading to vanishing $\tilde{J}_2^{ij}$ contribution.  Hence 
$\tilde{F}_{PV} = \tilde{J}_1 \, \vec{v}_{ep} \cdot \vec{S}_{ep} $ and 
$\tilde{F}_{PC} = \tilde{J}_1 \, \vec{S}_{ep}^2 $.
On the other hand, in case (1)
\begin{eqnarray}
&& 
\tilde{F}_{PV} = \tilde{J} \, \vec{v}_{ep} \cdot \vec{S}_{ep}
\,, \hspace{0.5cm}
\tilde{F}_{PC} = \tilde{J} \, \vec{S}_{ep}^2
\,,
\\ &&
\tilde{J} =  \int dE_1 \int dE_2 \delta ( E_1 + E_2 - \epsilon_{ep})
\nonumber \\ &&
\cdot
\int_{-\pi}^{\pi} d\theta_1 |\sin \theta_1| \int_{-\pi}^{\pi} d\theta_2 |\sin \theta_2| 
p_1 p_2  \, U_0(x) W_0(y)
\nonumber \\ &&
\hspace*{-0.5cm}
\cdot \left( E_1 E_2 -  p_1 p_2( \cos (\theta_1 - \theta_2) -2 \cos \theta_1 \cos \theta_2 )
- \delta_M  \frac{m_1 m_2 }{4} 
\right)
\,,
\nonumber \\ &&
\end{eqnarray}
where both $ \vec{v}_{ep} \,, \vec{S}_{ep}$ are parallel to the first excitation axis $\propto \vec{k}_1$.
The quantity ${\cal R}_{ij} $ of eq.(\ref {rate formula 0}) is thus
\begin{eqnarray}
&& 
{\cal R}_{ij} = \left( E_1 E_2 -  p_1 p_2( \cos (\theta_1 - \theta_2) -2 \cos \theta_1 \cos \theta_2 ) \right) |b_{ij}^2 |
\nonumber \\ &&
- \delta_M  \frac{m_1 m_2 }{8} \Re ( b_{ij}^2)
\,.
\label {r ij}
\end{eqnarray}
RENP rates are numerically
\begin{eqnarray}
&&
\Gamma_{{\rm RENP}} = 2 C|\vec{S}_{ep}|^2 |\chi|^2  \sum_{ij} {\cal F}_{ij} =
3.17 \times 10^{9}  \,{\rm sec}^{-1} {\rm eV}^{-5} 
\nonumber \\ &&
\cdot
|\chi|^2 \frac{ \sum_{ij} {\cal F}_{ij} }{{\rm eV}^5}
\frac{ 57.3\, {\rm sec}^{-1}}{\gamma_{ep} } \, \frac{ I_t}{ 10 {\rm mW cm}^{-2} }
\nonumber \\ &&
\cdot
(\frac{R }{0.5 {\rm cm} } )^4 (\frac{ L}{ {\rm cm}} )^2
(\frac{ n}{1.4 \times 10^{19} {\rm cm}^{-3} } )^2
\,.
\label {renp rate}
\end{eqnarray}

We turn to parity violating magnetization.
Amount of generated magnetization by neutrino pair emission is parity violating amplitude
multiplied by magnetic moment in the state $|f\rangle$, hence its magnitude is
\begin{eqnarray}
&&
M = n \mu_B \langle f| 
\hat{k}\cdot(\vec{L} + 2 \vec{S})
| f \rangle C |\chi|^2 \tilde{F}_{PV}
\nonumber \\ &&
= n g \mu_B \langle f| 
\hat{k}\cdot\vec{S}
| f \rangle C |\chi|^2 \tilde{F}_{PV}
\,.
\label {magnetization 1}
\end{eqnarray}
We use the relevant component of g-factor,
 $g_{\perp}$ of $ 5.9 $ for Er$^{3+}$ in the 
first excited J-manifold $^4$I$_{13/2}\,\Gamma_{56}$ of trivalent Er ion 
\cite{g-factor of 3+er}.
The number density of Er ions in 0.1 \% Er doped crystal (YLF)  is
$1.4 \times 10^{19} $cm$^{-3}$, which we shall use as a reference value.
We further introduce a conversion factor to magnetic field
from magnetic moment, $\xi$, since magnetization is measured outside the target, which gives
$\mu_{{\rm eff}} = \xi g \mu_B$ with
\begin{eqnarray}
&&
 \mu_{{\rm eff}}n = \, 0.29 {\rm G}\, \frac{\xi}{10^{-2}} 
\frac{n}{1.4 \times 10^{19}{\rm cm}^{-3}} 
\,,
\end{eqnarray}
$\xi $ is related to how near the target one measures magnetic field.
To convert the generated magnetic moment $\vec{\mu}$
 to the magnetic field caused by these moments, we use the formula of magnetic field
due to magnetic moment $\vec{\mu}$,
\begin{eqnarray}
&&
\vec{B}(\vec{r}) = - \int d^3r' n(\vec{r}')
 \left( \vec{\mu}\, \vec{\nabla}^2 \frac{1}{ |\vec{r}-\vec{r}' |}
- \vec{\nabla} ( \vec{\mu}\cdot\vec{\nabla}) \frac{1}{  |\vec{r}-\vec{r}' |}
\right)
\,,
\nonumber \\ &&
\end{eqnarray}
to be integrated over the target cylinder region ($ n(\vec{r}')$ is the number density
of RENP affected ions).
When the measurement site is chosen to be close to the target region,
the magnetic field generated by RENP may be  effectively expressed as $\xi \mu n $, $\xi$
being a fraction of RENP affected ions $\approx 0.01 $ for reference parameters.
Actual value of $\xi$ to be taken in magnetization depends on
how experiment is done.

Magnetization generation rate caused by neutrino pair emission is calculated 
using eq.(\ref{magnetization 1}) 
with $\langle \hat{k}\cdot \vec{S} \rangle = \cos \theta_s$
(signal photon directional factor) and replacement $\tilde{S }_{ep}^2 $ in rate by  $\vec{v} \cdot\tilde{S }_{ep} $
for magnetization.
 Weight of individual pairs is also changed to $b_{ij} c_{ij}$ for parity-odd magnetization.
These lead to 
\begin{eqnarray}
&&
2 C |\chi|^2 \xi \mu n\, \vec{S}_{ep}\cdot\vec{v}_{ep} \cos \theta_s \,\sum_{ij} {\cal G}_{ij}
\,, 
\\ &&
\hspace*{-0.5cm}
{\cal G}_{ij} = 
\int dE_1 \int dE_2 \,  \delta (E_1 + E_2 - \epsilon_{ep}) \,
\sqrt{E_1^2 - m_i^2} \sqrt{E_2^2 - m_j^2}  
\nonumber \\ &&
\hspace*{-0.5cm}
\cdot \int_{-\pi}^{\pi}  d\theta_1 
\int_{-\pi}^{\pi}  d\theta_2 \, | \sin\theta_1 \sin \theta_2 |
 \, U_0(\frac{K_{z} L}{2}) W_0(K_{\perp} R) \, {\cal M}_{ij} 
\,, 
\\ &&
\hspace*{-0.5cm}
{\cal M}_{ij} = 
\left(  E_1 E_2 - p_1 p_2 ( \cos (\theta_1 - \theta_2) 
- 2 \cos\theta_1\cos\theta_2 \right)  \Re( b_{ij} c^*_{ij})
\nonumber \\ &&
- \delta_M \frac{m_i m_j}{8} \Re (b_{ij} c_{ij})
\,.
\end{eqnarray}
The angular cosine $\cos \theta_s$ arises from 
$\vec{k}\cdot\vec{S}_{ep}/k \propto \cos \theta_s$.
Using $|\vec{v}_{ep} | |\vec{S}_{ep}| = 0.0157 \times 2.49 \times 10^{-6}/3.47 = 1.1 \times 10^{-8}$,
calculated magnetization at neutrino pair emission is numerically
\begin{eqnarray}
&&
4.22  \times 10^{4}
\, {\rm  G\, sec}^{-1}\, |\chi|^2 \,\frac{\xi }{10^{-2}} \,
\frac{I_t}{ 10 {\rm mW cm}^{-2}} \frac{ \sum_{ij} {\cal G}_{ij}}{{\rm eV}^5}
 \cos \theta_s 
\nonumber \\ &&
\cdot
\frac{\vec{S}_{ep}\cdot\vec{v}_{ep} }{|\vec{v}_{ep}|  | \vec{S}_{ep} | } 
( \frac{ n}{ 1.4 \times 10^{19} {\rm cm}^{-3}})^3 
(\frac{R }{1 {\rm cm}})^4 (\frac{ L}{2 {\rm cm}})^2
\,.
\label {2nu integral pv2}
\end{eqnarray}
The number in front is for 
1000/57 msec of $1/\gamma_{ep}$, 
and if the value $\gamma_{ep} = 2\pi \times $ 10 MHz is taken,
it is changed to $1 \times 10^{-3}$.
A typical value for
computed $ \sum_{ij} {\cal G}_{ij}$ is of order $10^{-8}$ eV$^5$,
implying generated magnetization of order,
$ 10^5 |\chi|^2 $ nG sec$^{-1}$
$L = 2R = 2 $cm. 
See below on $|\chi|^2$.

We work out results of magnetization angular
distribution for case(1);
 the case of c-axis parallel to excitation laser wave vector $\vec{k}_1$.
Rate is proportional to ${\cal F}_{ij} \propto b_{ij}^2$,
while magnetization is to $\cos \theta_s\, {\cal G}_{ij}\, \propto b_{ij} c_{ij}$,
assuming CP conservation of all vanishing phase $\delta = \alpha = \beta =0$.
Different weights in rate and magnetization distributions in six neutrino-pair
thresholds are  calculable from neutrino oscillation data
\cite{pdg}, and they are
listed in the following table, Table 4, assuming CP conservation case described by
real number $b_{ij}, c_{ij}$.

\vspace{0.5cm}
\hspace*{-1.5cm}
\begin{tabular}{c|c|c|c|c|c|c} \hline 
$(ij)$ &(11) & (12) & (22) & (13) & (23) & (33)\\ \hline \hline
${\rm rate} \propto b_{ij}^2$ & 
0.0311 & 0.405 & 0.0401  & 0.0325 & 0.0144 & 0.227 \\ \hline
${\rm magnetization} \propto b_{ij} c_{ij}$ & 0.115&0.405 & 
0.1356 & 0.0325 & 0.0144 & 0.454
 \\ \hline
\end{tabular}
\begin{center}
Table 4 Relative weights of neutrino-pair thresholds as determined
by neutrino oscillation data, assuming CP conservation.
\end{center}

\hspace*{-1.5cm}
\begin{figure*}[htbp]
 \begin{center}
 \epsfxsize=0.4\textwidth
 \centerline{\includegraphics{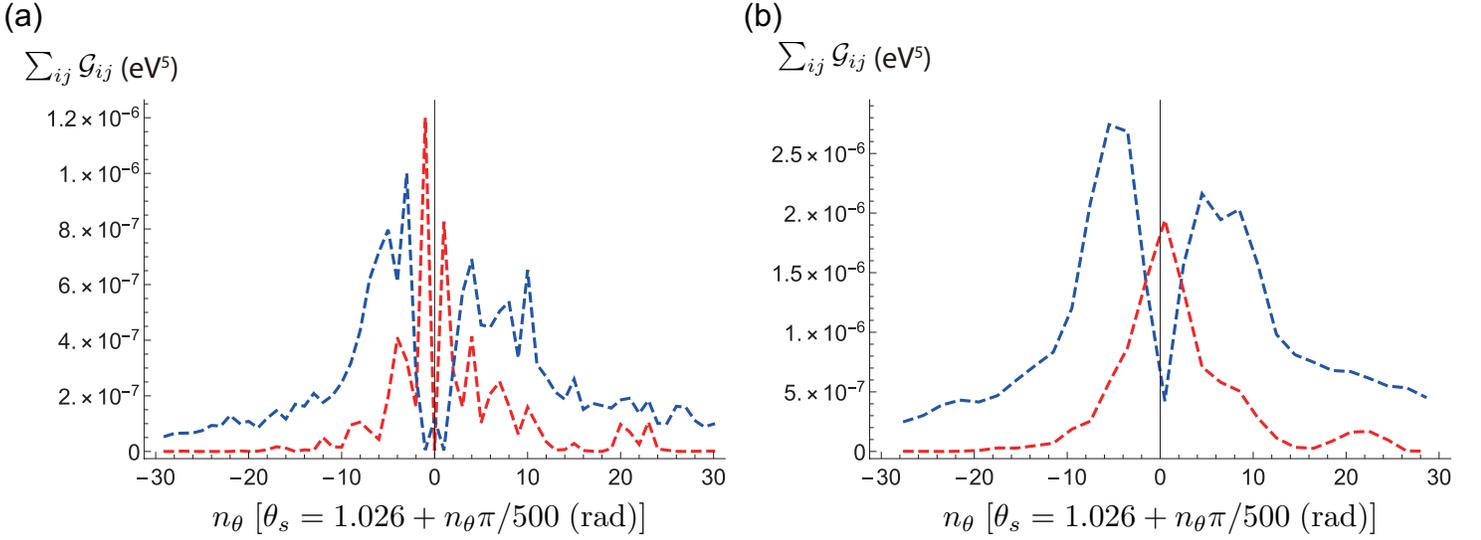}} \hspace*{\fill}\vspace{-10cm}
   \caption{
(a) 
Angular distributions (relatively rescaled for easier comparison
of different size results)
of $\sum_{ij} {\cal G}_{ij}(\theta_s)$ using two sizes of crystals:
small $(R/cm\,, L/cm\,, V/cm^3) = (0.25\,, 0.5\,, 0.0982)$  (in blue)
and large $(1\,, 2\,,6.28) $ (in red) crystal 
sizes  for Dirac NH (normal hierarchy) of smallest mass $m_1=10$ meV.
The angular region covers 21.6$^{\circ}$ around the expected
$59^{\circ}$ signal direction.
(b) Smoothed $\sum_{ij} {\cal G}_{ij}$ corresponding to (a).
Values at consecutive four angular points $\theta_s$ covering 1.44$^{\circ}$
 are added with the half angle overlap to adjacent regions as in
Fig(\ref{excitation and de-excitation}a).
}
   \label {dnh large and small crystals}
 \end{center} 
\end{figure*}

We now address the postponed question of how the size or volume of crystal
targets affects the angular distribution, and wish to estimate 
an optimized target crystal size for experiments. 
For definiteness we take the trivalent Er scheme 
identified by eq.(\ref{er3+ scheme 1}) $\sim $ (\ref{er3+ scheme 4}),
and focus on excited region of  either cylinder or disc shape whose
boundary lies near the aspect ratio of $R/L = 1/2$.
It is always easy to experimentally study  excited region of smaller aspect
ratio, $2R< L$ by laser focusing,
hence we concentrate to the aspect ratio 1/2, $V = \pi L^3/4$,  for subsequent study.
We may classify small, medium and large crystals with varying volumes,
taken here in a range $ L = 0.5 \sim 5 $cm,
since much smaller and much larger crystal sizes have problems either 
of   bad angular resolution or of crystal growth in actual experiments.
Macro-coherent rate (magnetization) is proportional to 
angular function, $V^2 \sum_{ij} {\cal F}_{ij} (\theta_s)$
($V^2 \cos \theta_s\, \sum_{ij} {\cal G}_{ij} (\theta_s)$)
given by a unit of ${\rm eV}^{5} {\rm cm}^6$.

We first plot magnetization strength given by
$\sum_{ij} {\cal G}_{ij} (\theta_s)$ in Fig(\ref {dnh large and small crystals}),
in two cases of small and intermediate crystal sizes, $L = 0.5$ and $ 2$ cm's.
The unit of angular resolution in this figure is 0.36$^{\circ}$ covering
a region of 21.6$^{\circ}$  centered at 59$^{\circ}$.
The resolution 0.36$^{\circ}$
seems smaller than experimentally attainable angular resolution,
hence it is better for comparison with experimentally obtained data to
theoretically average over discretized angles by a smoothing procedure.
We adopt here a sum over four adjacent angular points (1.44$^{\circ}$), taking
two overlapping angular 0.72$^{\circ}$  points  with  neighboring regions.
This gives result of Fig(\ref{dnh large and small crystals}b) corresponding to
result of Fig(\ref {dnh large and small crystals}a).
As expected, smoothed angular distribution does not show
complicated wiggling structures.
We confirmed that central peaking increases as the crystal size increases, to
5 cm $\sim$ 10 cm, approaching the infinite volume limit.

This kind of smoothing should be incorporated in 
detailed design study of experiments,
but there are other factors such as spatial laser profile defining the
excited target region and angular resolution of measuring device of
magnetization, and consideration of these is outside the present scope
of work.
We shall therefore present numerical  results without smoothing in
subsequent results.
For simplicity we give results for $L=2R = 2$ cm size of crystal in subsequent figures.

\begin{figure*}[htbp]
 \begin{center}
 \epsfxsize=0.4\textwidth
 \centerline{\includegraphics{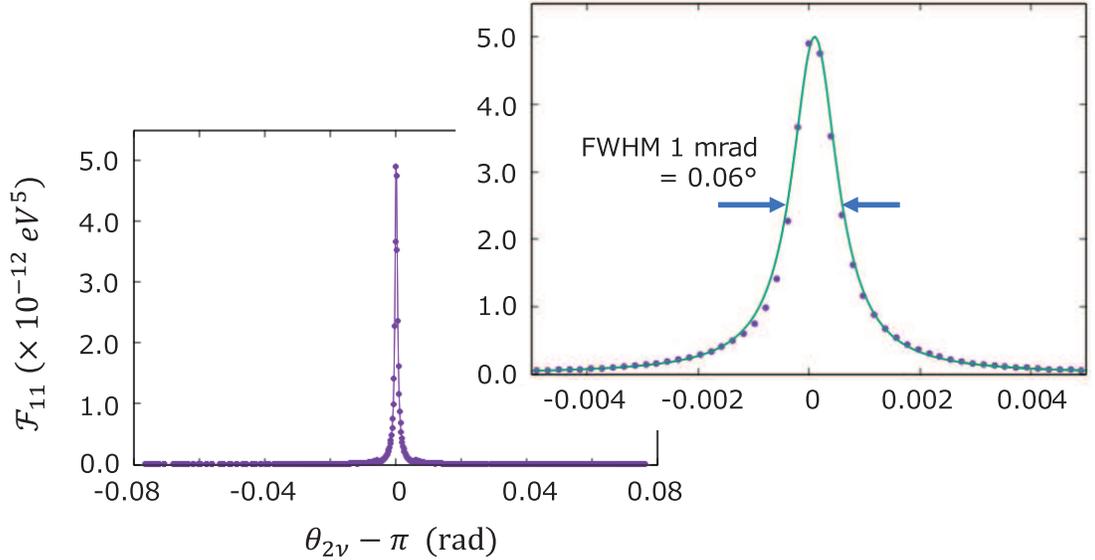}} \hspace*{\fill}
   \caption{
Opening angle ($\theta_{2\nu}= \theta_1-\theta_2 $) distribution of 
emitted neutrino pair of mass 10 meV for NH Dirac case,
contributing to PC term ${\cal F}_{PC}$ (in eV$^{5}$ unit), its values
plotted against $\theta_{2\nu} - \pi$.
}
   \label {nu-pair o-ang dist}
 \end{center} 
\end{figure*}

The first physics issue we would like to discuss is neutrino-pair
opening angle ($\theta_{2\nu} = \theta_1-\theta_2 $) distribution
given by $\tilde{ F}_{PC}$,
integrated over $\theta_i\,, i=1,2$ with the constraint of fixed
$\theta_{2\nu}  $ around $\pi$, which is illustrated in Fig(\ref{nu-pair o-ang dist}).
It shows a remarkable back-to-back configuration, consistent with the
momentum conservation.
We assumed a single neutrino-pair of  mass as small as 10 meV.

\begin{figure*}[htbp]
 \begin{center}
 \centerline{\includegraphics{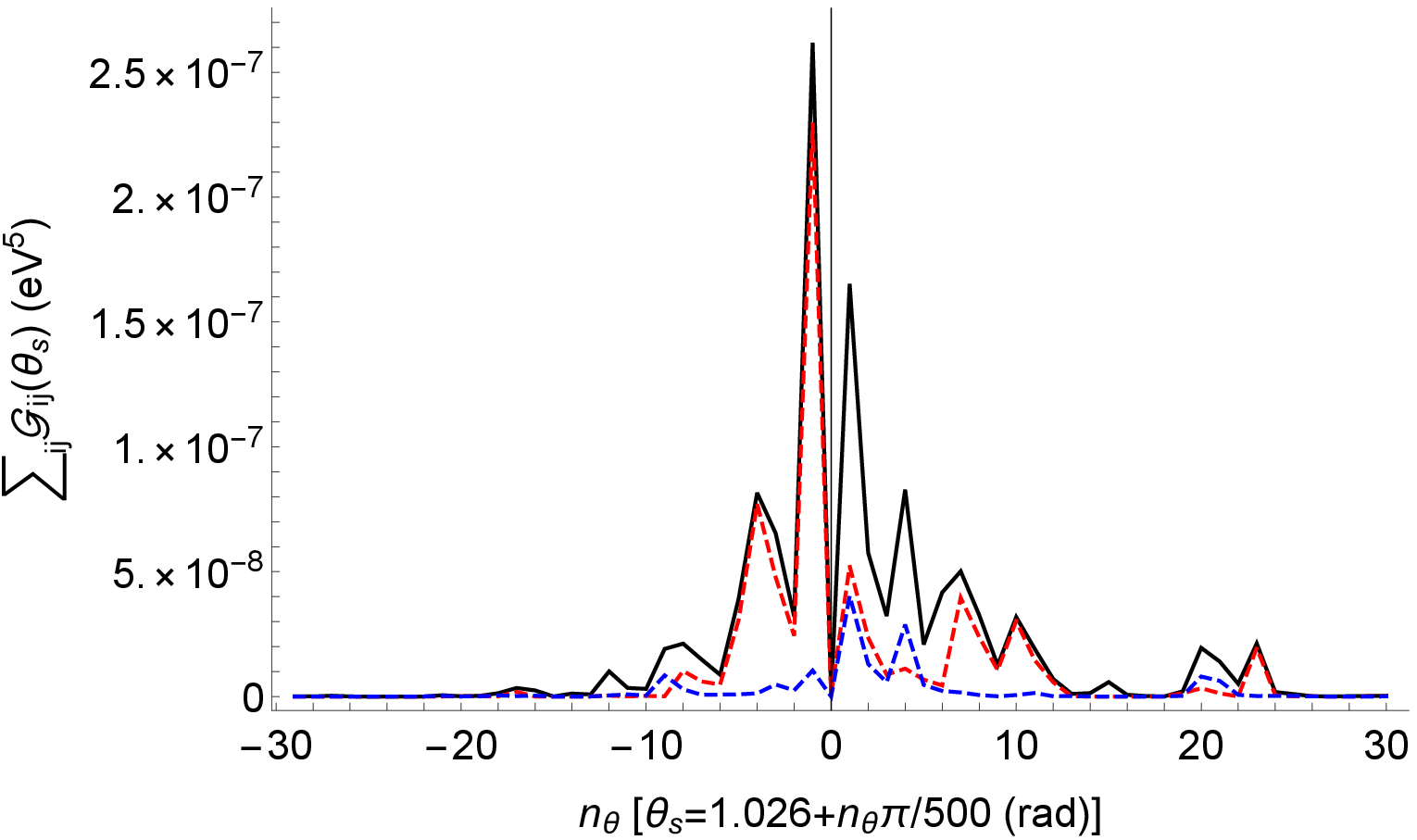}} \hspace*{\fill}
   \caption{
Decomposition into some individual neutrino-pair contributions
for $m_1 = 10$ meV NH Dirac case:
totally summed over six pairs distribution$ \sum_{ij} {\cal G}_{ij}$ in joined black, 
(12) pair contribution $ {\cal G}_{12}$ in joined dashed red,
and (33) pair contribution $ {\cal G}_{33}$ in joined dashed blue, for
$L=2R=2$ cm crystal size, taking
angular resolution $\pi/500$, covering a range 21.6$^{\circ}$.
}
   \label {pair decomposition}
 \end{center} 
\end{figure*}

Decomposition into individual $(ij)$ neutrino-pair contribution may 
be illuminating, although the task is not rewarded in experimental analysis.
Result is illustrated in Fig(\ref {pair decomposition}) showing different
angular structures for different pair contributions.
Note weight factors given in $ b_{ij}c_{ij}$ row of Table 4, showing two
largest (12) and (33) pair contributions.

\begin{figure*}[htbp]
 \begin{center}
 \epsfxsize=1.0\textwidth
 \centerline{\epsfbox{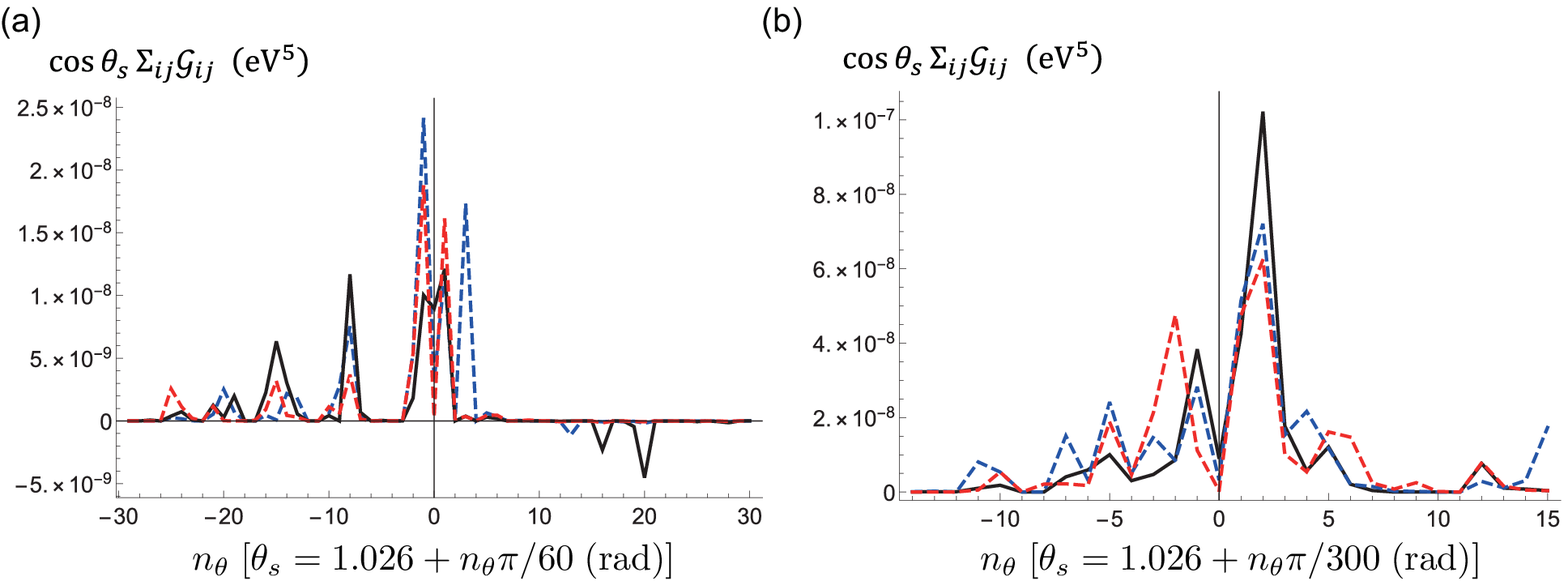}} \hspace*{\fill}\vspace*{1cm}
   \caption{ 
Magnetization angular distribution  $ \cos \theta_s \sum_{ij} {\cal G}_{ij}(\theta_s)$ in $ {\rm eV}^5$ unit: NH (normal hierarchically ordered) 
Dirac cases of smallest mass 10 meV 
in joined dashed red,  
30 meV in joined dashed blue,
 and 50 meV in joined black, for $L=2R=2$ cm crystal size, taking
angular resolution $\pi/60$ in (a), which  covers
the entire range 180$^{\circ}$, with $\theta_s = 1.026 $ at the center, and
enlarged central region by a factor 5 in (b).
Note that peak structures differ for these two different angular resolutions,
as expected. Smoothing procedure should thus be performed prior to
experimental comparison.
Absolute unit of generated magnetization rate to be multiplied is
$\sim 4.2 \times 10^4 |\chi|^2$ G sec$^{-1}$, as given 
in eq.(\ref{2nu integral pv2}),
leading to typical values, $\approx 4 \times 10^5 |\chi|^2 $ nG sec$^{-1}$.
}
   \label {er magnetization}
 \end{center} 
\end{figure*}

Magnetization angular distributions given by 
$\cos \theta_s \sum_{ij} {\cal G}_{ij}(\theta_s)$
are illustrated for Dirac NH cases of three smallest neutrino masses,
10, 30, and 50 meV's in 
Fig(\ref{er magnetization}a) and  Fig(\ref{er magnetization}b).
These mass differences are expected to be distinguishable in experimental analysis.

\begin{figure*}[htbp]
 \begin{center}
 \centerline{\includegraphics{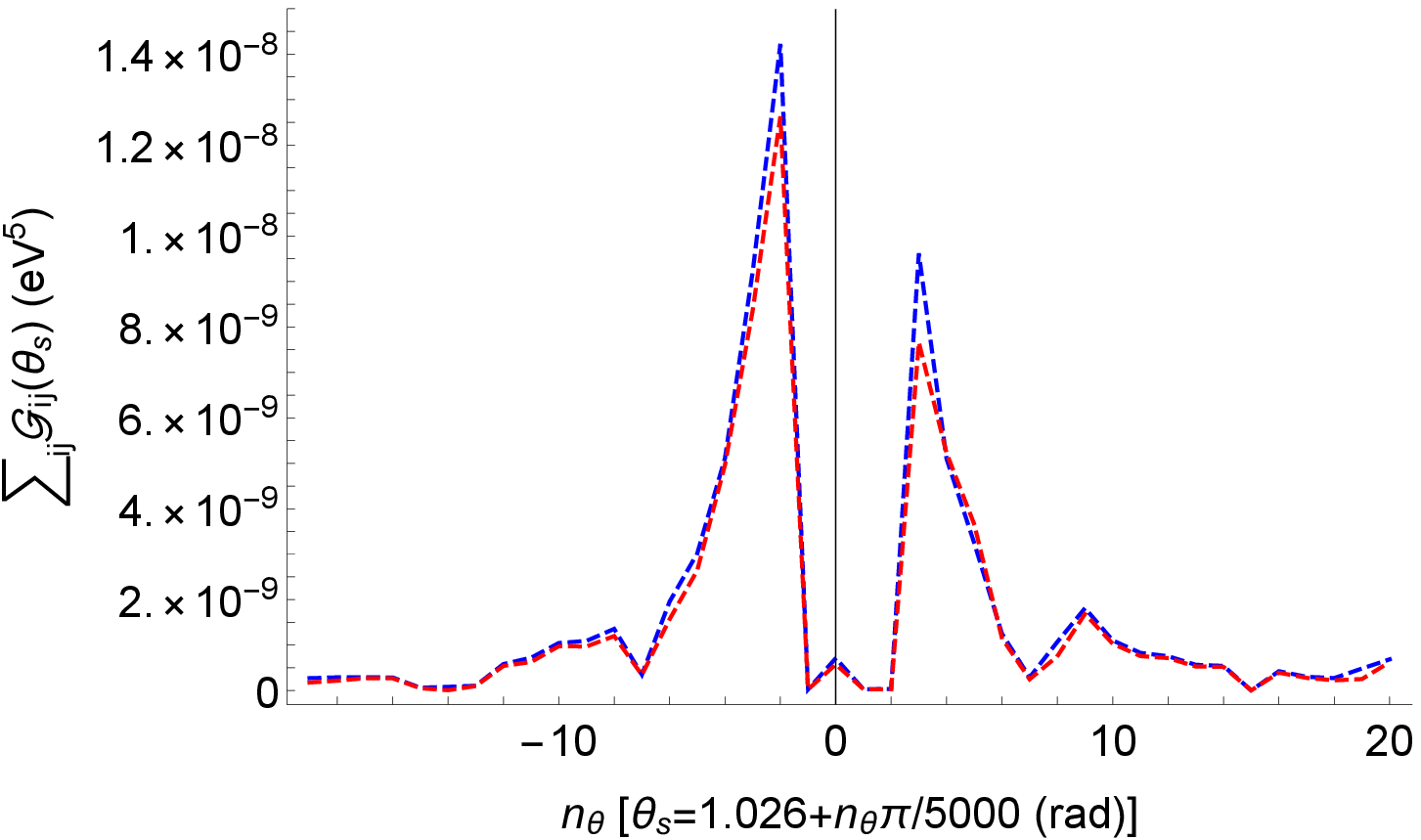}} \hspace*{\fill}
   \caption{ 
Majorana/Dirac distinction in terms of angular distribution of 
$ \sum_{ij} {\cal G}_{ij}(\cos \theta_s)$
 for NH smallest mass 600 meV case:
Dirac case in joined blue and Majorana case in red.
$L=2R =2$ cm crystal size, taking
angular resolution $\pi/5000$ covering 1.44$^{\circ}$.
Ratio of summed values is 1.11.
}
   \label {md distinction}
 \end{center} 
\end{figure*}

Majorana-Dirac distinction in terms of angular distribution
 is not easy as shown in Fig(\ref{md distinction}), which gives
Dirac and Majorana functions of $ \sum_{ij} {\cal G}_{ij}(\theta_s) $, 
taking a  NH smallest mass as large as 600 meV.
Ratio of angular point sums is 1.06 for $m_1= $500 meV and
1.11 for 600 meV (Dirac values larger than Majorana ones).
In order to detect Majorana/Dirac distinction for smaller neutrino masses,
it is better to 
use hyperfine levels of odd Er isotopes along with magnetic field,
which also requires a joint microwave spectroscopy.
This subject is beyond the present scope of work.

\begin{figure*}[htbp]
 \begin{center}
 \epsfxsize=0.8\textwidth
 \centerline{\epsfbox{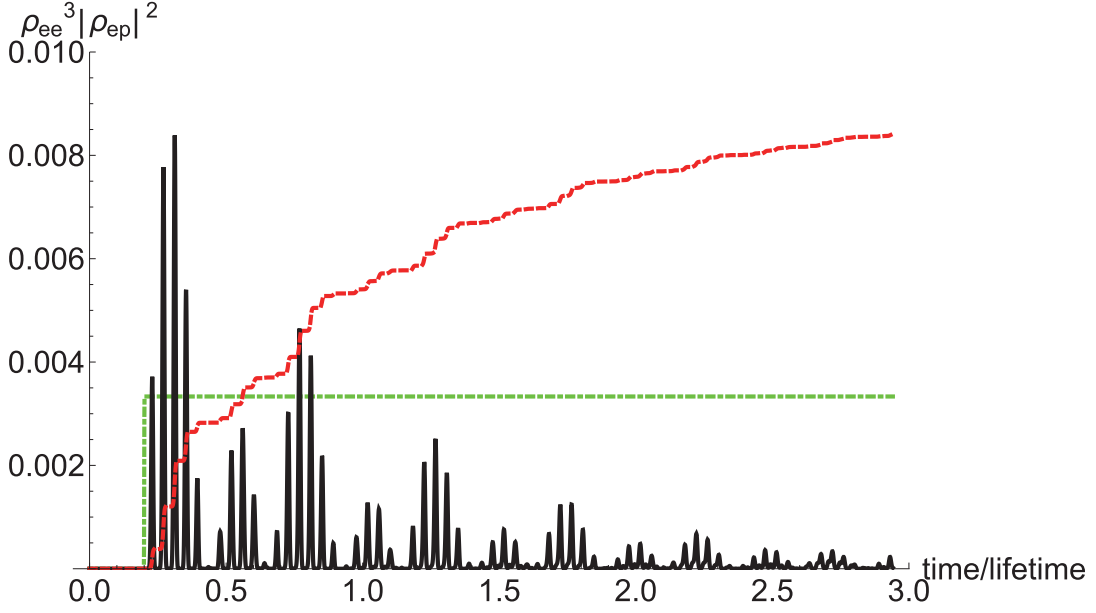}} \hspace*{\fill}\vspace*{1cm}
   \caption{
Time structure of magnetization generation rates  $\propto \rho_{ee}^3 |\rho_{ep}|^2$
(not to be confused with RENP event rate $\propto \rho_{ee}^2 |\rho_{ep}|^2$)
in solid black, 
when trigger is irradiated at 0.2 $\times $ lifetime of state $^2$H$_{11/2}$ with
time lag after irradiation of two-photon excitation lasers.
Also shown  is time profile of accumulated magnetization (ten times values of
directly integrated result) in dashed red, which remain in crystals within
this time scale.
The same set of parameters as in Fig(\ref{obe 4-level 2}) is used here,
with the green curve indicating trigger laser time profile.
}
   \label {renp t-profile}
 \end{center} 
\end{figure*}

Let us discuss the remaining important factor $|\chi|^2$.
We assume CW (continuous wave) laser operation, and CW
trigger laser is  irradiated with a time lag after two-photon excitation.
RENP events start to occur at trigger irradiation, and generated  magnetization
remains after events till spin relaxation time comes.
Time evolution of population $\rho_{ee}$ in the upper excited state $| e\rangle$
and coherence $\rho_{ep}$ between levels of neutrino-pair emission
is described by four-level optical Bloch equation, as given in Appendix.
The coherence $\rho_{eg}$ and population $\rho_{ee}$ 
develop prior to trigger irradiation, and 
the coherence is partially transfered to $\rho_{ep}$ after trigger irradiation.
Time dependent $|\chi|^2$ is identified to be $|\rho_{ee} \rho_{ep} |^2 $.
Generation rates of magnetization are proportional to 
$n |\chi|^2 \propto \rho_{ee}^3 | \rho_{ep}|^2$,
which has a time structure, as illustrated in Fig(\ref{renp t-profile}).
Note that event rate has a different power dependence $\rho_{ee}^2 | \rho_{ep}|^2$.
In our example of numerical computation, generated value $\rho_{ee}^3 | \rho_{ep}|^2$
becomes of order, $10^{-2} \sim 10^{-3}$.
It is important to note that this large value is realized for weak intensity $I_1, I_2, I_t$
of three CW lasers, which however requires a small relaxation rate $\gamma_{ep}$.
The largest RENP event occurs right after the trigger irradiation, accompanied by
side events, which is somewhat similar to ringing
phenomena of super-radiance events \cite{dicke 2}.
As stressed, signal of RENP events persists as magnetization till
spin relaxation time, hence time-accumulated  events  show increasing
 profile of magnetization with time as in Fig(\ref{renp t-profile}).
We confirmed that laser turn off leads to gradual decrease of $\rho_{ee}^3 | \rho_{ep}|^2 $.
In actual experiments time-integrated magnetization till spin relaxation time
is expected to be a readily observable quantity.

Our calculated magnetization rate per second is applicable at cycle end 
of three laser operation, and cycles are repeated with
a cycle repetition time to be taken ideally of order spin relaxation time.
Thus, the quoted rate per second so far given should be multiplied by
the number of repetition times per second.
If the repetition cycle is 100 per second,
the effective rate per second would be increased by 100.
It would be nice if one can make dedicated simulations including both
spin relaxation and coherence development at the same time.
For this simulation data of spin relaxation is needed.
This is left to future works.

\section 
{QED  background}

Major obstacle of QED background is the problem of extinction of initial ions 
in state $|e\rangle $ for RENP.
We shall discuss amplified processes first and at the end of this section we discuss effect 
of spontaneous electric dipole transition to the ground state.

We consider potentially dominant amplified QED event of
two photon emission,
$|e\rangle \rightarrow | g \rangle + \gamma_s + \gamma_b$ via $|p\rangle$,
with extra background photon  $\gamma_b$ of energy $\epsilon_{ep}$.
Three vectors, $\vec{k}_1\,, \vec{k}_2\,, - \vec{k}_s$, are arranged to be in
the same plane, hence four momentum vectors, 
$\vec{k}_1\,, \vec{k}_2\,, - \vec{k}_s\,, - \vec{k}_b$,
form closed quadrangle to leading order of momentum conservation.
One can directly prove that there is no extra photon vector $\vec{k}_b$
satisfying the quadrangle relation, using parameter sets of proposed scheme.
Still, the improved method of plane-wave summation in this work gives
a finite contribution.
Our estimate of finite size effect gives extinction rate per ion
of order $3 \times 10^{-4} $, taking inhomogeneous width $\gamma_e $ to be 1 MHz
(effect $\propto \gamma_e^{-2}$).
This rate  is sufficiently small to ignore the process.

The next sub-dominant contribution is amplified three-photon emission,
$|e\rangle \rightarrow | g \rangle + \gamma_s + \gamma_b + \gamma_b'$
with $\gamma_b + \gamma_{b'} = \epsilon_{ep}$.
In order to suppress this process, we resort to symmetry and apply a weak magnetic
field of order 100 G to distinguish  by irradiated laser frequency 
two Zeeman states of separated doublet components.
Zeeman split levels are given in terms of states 
of definite time reversal quantum numbers $\pm$, 
\begin{eqnarray}
&&
\hspace*{-0.3cm}
|J, \pm J_z \rangle = \frac{1}{\sqrt{2}} (|J, |J_z|, + \rangle \pm |J, |J_z|, - \rangle )
\,, \hspace{0.5cm} J_z > 0
\,.
\end{eqnarray}
Optical transitions between two states of $|e\rangle, |p\rangle$
given by $|J, \pm J_z \rangle $
and $|J', \mp J'_z \rangle $, proceed via either single photon T-odd operator $V_o$ or
two-photon T-even operator $V_e$ via paths of different energy,
$| J, \pm J_z \rangle \rightarrow | J', \mp J'_z \rangle$ for T-odd case,
$| J, \pm J_z \rangle \rightarrow | J', \pm J'_z \rangle$ for T-even case.
This situation occurs due to cancellation of two terms,
for instance between $|J, J_z, \pm \rangle \rightarrow |J', J'_z, \pm \rangle$
for T-odd case.
One can thus single out T-odd transition by laser frequency,
rejecting two photon emission.
On the other hand,
when Kramers degeneracy exists in the zero field,
both T-even and T-odd transitions occur since lasers cannot distinguish two modes of
transitions.

This leaves amplified four-photon emission,
$|e\rangle \rightarrow | g \rangle + \gamma_s + \gamma_1 + \gamma_2 + \gamma_3$, 
of QED origin as the main extinction process.
Estimate of  total rate of this process is as follows.
Except a common factor related to signal photon part,
the major difference arises from coupling factors and phase space integration
either over two-body or three-body states.
Couping factors of QED background is of order, (dipole$^2$/3)$^3 \approx
10^{-39}$eV$^{-6}$, while  RENP  coupling is of order $G_F^2 \approx 10^{-46}$eV$^{-4}$.
Three-body phase space of background QED photons have
\begin{eqnarray}
&&
(\frac{2}{3})^3 \int \frac{d\omega_1 d\omega_2 d\omega_3}{(2\pi)^5} \delta (\sum_i \omega_i
- \epsilon_{ep}) \frac{\omega_1^3 \omega_2^3 \omega_3^3}{ \Delta_{1}^2 \Delta_2^2}
\nonumber \\ &&
= (\frac{2}{3})^3 \frac{\epsilon_{ep}^{11}}{ (2\pi)^5\bar{\Delta}_{1}^2 \bar{\Delta}_2^2}
\frac{1}{184800}
\,,
\end{eqnarray}
where energy denominators $\Delta_i\,, i = 1,2$ that appear in perturbation theory
 are linear combinations of level energy differences $\epsilon_{ab}$ and
one of emitted photon energies $\omega_i\,, i = 1,2,3$,
with $\bar{\Delta}_{i}$ its  appropriate average value,
while two-neutrino phase space has
\begin{eqnarray}
&&
(4\pi)^2 \int \frac{dE_1 dE_2 }{(2\pi)^5} \delta (\sum_i E_i
- \epsilon_{ep}) = \frac{2}{15 (2\pi)^3} \epsilon_{ep}^5
\,.
\end{eqnarray}
Thus, the ratio of three- to two-body phase space is of order,
$ 3 \times 10^{-7} (\epsilon_{ep}^6/\bar{\Delta}_{1}^2 \bar{\Delta}_2^2 )$,
which gives roughly comparable magnitude of amplified QED background to
RENP events.
Even if the four-photon process depletes to some extent the excited state $|e\rangle$,
parity violating magnetization can single out RENP process.

Finally, we discuss how magnetization may be measured under a 
spontaneous electric dipole transition from state $|e \rangle$ to the ground J-manifold of lifetime $1/518$ sec (from Table 1).
Despite of small branching ratio of order $10^{-12}$
(proportional to laser power, taken here 10 mW cm$^{-2}$, and other factors, hence controllable to some extent)
RENP events occur with reasonably large total event number
(assuming reference target ion density and target size)
\begin{eqnarray}
&&
\Gamma_{{\rm RENP}} \tau_e = 0.19 |\chi|^2 \frac{\sum_{ij} {\cal F}_{ij}}{10^{-8}} \frac{I_t}{10 {\rm mW cm}^{-2}} 
\,,
\end{eqnarray}
during lifetime of $|e \rangle$.
Accumulated magnetization is measurable by taking sufficient measurement time
with a fast repetition cycle of laser excitation.
Although leakage due to spontaneous decay occurs to different time reversal state, this is not a problem since CW two-excitation to $|e\rangle $ followed by another spontaneous emission takes the excited state back the original path.

\section 
{Summary and Outlook}

Even if the experimentally confirmed macro-coherence amplification \cite{psr exp 1},
\cite{psr exp 2}, \cite{psr exp 3}, works,
the major problem for a successful  RENP measurement of process
$|e \rangle  \rightarrow | g \rangle + \gamma_s+ \nu\bar{\nu} $
from an excited state $|e\rangle$  is 
how to reject similarly amplified QED backgrounds.
We proposed in this paper a method of overcoming this difficulty by
measuring parity violating observable present in
weak interaction and absent in QED.
The proposed magnetization along the signal photon direction
is parity violating, and emerges with reasonably large magnitudes,
if one adopts a clever way of choosing excitation and trigger  directions
and their photon energies.
The best candidate target we have found so far is  trivalent lanthanoid ion of
odd number of 4f electrons: Er$^{3+}$ doped in
host crystals such as YLF. Host crystals should
provide to the ion a sharp radiative width and a large spin relaxation time.
Moreover, the Kramers degeneracy of odd number 4f systems
such as Er$^{3+}$ brings about an experimental bonus.
Excitation at $|g\rangle \rightarrow |e\rangle $ is 
done by T(time reversal)-even two-photon  process to
imprint a phase to target ions, while macro-coherence generation 
between states of neutrino pair emission
is achieved via single photon T-odd trigger laser irradiation with a time lag.
Numerical simulations using four-level optical Bloch equation supports
a sizable generation of coherence and population among relevant states.

Calculation of magnetization suffers from uncertainty,
most notably due to a lack of optical experimental data 
of separated magnetic and electric dipole transitions between two relevant states
of neutrino pair emission.

Coexistence of magnetic and electric dipole transitions is a key element
for parity-violating magnetization measurement.
Crystal field of surrounding host ions introduces an environmental parity violation,
a necessary ingredient for a successful measurement of
RENP originated magnetization.
Generated magnetization may become of order $10^{2} $ nG sec$^{-1}$
for 0.1\% concentration of trivalent Er ion, 
depending on coherence factor
generated at excitation.
Numerical analysis of RENP rate and magnetization have been carried out by
incorporating finite size effect of target crystal, which is important due to realistic
crystal sizes of order 1 cm hopefully used in experiments.

If the angular resolution of magnetization measurement is reasonably good,
of order a few degrees,
magnetization has a good sensitivity to
the smallest neutrino mass down to of order 20 meV. 
Majorana/Dirac distinction is difficult from measurement of angular distribution
of magnetization in the proposed Er$^{3+}$ scheme.
One possibility for improvement is to use another crystal such as Ho$^{3+}:$YLF.

We did not study sensitivity to CP violation parameters, 
but this may open a new possibility of Majorana/Dirac determination assuming
presence of CPV phases, because finite CPV phases
may enhance Majorana/Dirac difference term
proportional to $m_i m_j \Re(b_{ij} c_{ij} ) $.
There are two ways to determine CP violation parameters in the mixing matrix:
one is to incorporate CP phases in time reversal even quantities such as rate and
magnetization rate, which is a straightforward extension of our formalism here.
This gives even functions of CP phases.
The other is to measure time reversal violation directly
and, for this purpose, to calculate time reversal odd measurable which
is given by odd function of CP phases.
These studies are left to future work.

A rich variety of angular structures resulting from
incorporation of finite size effect
as seen in our figures
is not really what we can precisely measure
in actual experiments, since detector angular resolutions, in particular
for magnetization measurement using SQUID, should be taken into account.
We may expect to observe smoothed out distributions,
but details of the smooth-out should be worked out
taking into account realistic angular resolution of adopted detector.
It is rather better to first set up experimental goals for
neutrino parameters, and to design and choose detector system
matching this objective by detailed simulations using
the methods developed in the present work.
In the meantime it is desirable to measure electric and magnetic dipole
transition rates between  two relevant ion states in crystals
such as Er$^{3+}:$YLF.


\section 
{Appendix: Optical Bloch equation for estimate of
population and coherence
 }

Before we present four-level optical Bloch equation
and its numerical simulations,
we give an intuitive picture of what may be realized in ideal situations.
Suppose that trigger laser irradiation is followed
by two-laser excitation after a time lag.
All lasers are assumed to be operated by CW (continuous wave) mode,
which effectively means that lasers are irradiated during the entire
lifetime scale.
One may take Rabi frequencies much
larger than radiative decay rates, $\Omega_{ab} = \sqrt{d_{ab}^2 +\mu_{ab}^2} E \gg \gamma_{ab}$,
using a reasonable range of laser intensity, $I_t = E^2 = B^2$.

CW steady state solution derived by analysis of
ladder-type three-level optical Bloch equation after various lifetimes
 has the following form of population and coherence, in the zero detuning limit,
 \cite{3level bloch}, \cite{obe my},
\begin{eqnarray}
&&
\left(
\begin{array}{c}
\rho_{ee}  \\
\rho_{gg}  \\
\rho_{qq}  \\
\rho_{eg}  \\
\rho_{eq}  \\  
\rho_{qg}  
\end{array}
\right)_0 =
\frac{ 1}{ \Omega_{eq}^2 + \Omega_{qg}^2}
\left(
\begin{array}{c}
\Omega_{qg}^2  \\
\Omega_{eq}^2  \\
0 \\
-\Omega_{eq}\Omega_{qg}   \\
0 \\  
0
\end{array}
\right) + O(\gamma_{ab})
\,.
\end{eqnarray}
It is thus possible to generate a maximal coherence, $|\rho_{eg}| = 1/2 $ between 
$|e \rangle $ and $|g \rangle $, with population ratio of $1:1: 0$ between
three levels $|e \rangle, | g\rangle, |q \rangle$ when $\Omega_{qg} = \Omega_{eq} $.
A nearly complete inverted population to $|e \rangle $ is also possible when
$\Omega_{qg} \gg \Omega_{eq}$, the inequality sometimes felt counter-intuitive.

Suppose that trigger laser matched to energy difference  of  $|g\rangle $ and $|p\rangle $
is irradiated with a time lag
when the steady state caused by two-photon excitation is being formed.
One can then effectively consider three level V-scheme in which two
lasers irradiated between $|g\rangle \,, | e\rangle$ are replaced by an effective single laser,
forming V-type laser irradiation along with trigger $|g\rangle \rightarrow | p\rangle$.
The steady-state solution of V-type is given by \cite{obe my}
\begin{eqnarray}
&&
\hspace*{-0.5cm}
\left(
\begin{array}{c}
\rho_{ee}  \\
\rho_{gg}  \\
\rho_{pp}  \\
\rho_{eg}  \\
\rho_{ep}  \\  
\rho_{pg}  
\end{array}
\right) =
\frac{ 1}{ 2(\Omega_{eg}^2 + \Omega_{pg}^2)}
\left(
\begin{array}{c}
\Omega_{eg}^2  \\
\Omega_{eg}^2 + \Omega_{pg}^2  \\
\Omega_{pg}^2 \\
0 \\
 \Omega_{eg}\Omega_{pg} \\  
0
\end{array}
\right) + O(\gamma_{ab})
\,.
\end{eqnarray}
One can thus generate a large coherence between states, $|p\rangle \,, | e\rangle$,
provided that Rabi frequencies, an effective
$ \Omega_{eg}$ and $\Omega_{pg}$, are of comparable magnitudes.
For instance, $\rho_{ep} = 1/4$ for $\Omega_{pg} = \Omega_{eg}$.
This is important, because neutrino pairs are emitted between these states
and large coherence is required for $\rho_{ep}$.

The picture here is a kind of coherence transfer from
$\rho_{eg}$ at excitation to $\rho_{ep}$ at trigger.
We would like to check whether this picture
is realized in four-level problem under three laser excitation advocated in the text.

The four-level optical Bloch equation 
of our interest concerns 16 component vector $ \vec{\sigma}_4$, 
and is given by a set of linear differential equations \cite{lindblad},
\begin{eqnarray}
&&
\frac{d}{dt} \vec{\sigma}_4 =  {\cal R}_4 \vec{\sigma}_4 
\,, 
\\ &&
\vec{\sigma}^T_4 = (\,\rho_{ee }\,, \rho_{gg}\,, \rho_{qq}\,,\rho_{pp}\,,
 e^{-i (\delta_{eq}+ \delta_{qg}+ \delta_{pg} )t} \rho_{ge}\,, 
\nonumber \\ &&
e^{i (\delta_{eq}+ \delta_{qg} + \delta_{pg}) t} \rho_{eg}\,, 
e^{- i\delta_{eq} t } \rho_{qe}\,, 
e^{ i\delta_{eq} t} \rho_{eq}\,, 
e^{i \delta_{qg} t} \rho_{qg }\,, 
\nonumber \\ &&
\hspace*{-0.5cm}
e^{- i \delta_{qg}t } \rho_{gq }\,, 
e^{i \delta_{pg} t} \rho_{pg }\,, e^{- i \delta_{pg}t } \rho_{gp }\,,
\rho_{pe}\,,  \rho_{ep}\,,
\rho_{pq }\,, \rho_{qp } \,)
\,,
\end{eqnarray}
where $16 \times 16 $ ${\cal R}_4$  matrix has the following form
in the case of no de-tuning of $\delta_{ab} =0$,
\begin{eqnarray}
&&
{\cal R}_4 =
\left(
\begin{array}{ccc}
A & C_1  & C_2 \\
D_1 & B_1    & E \\
D_2 &  F   & B_2
\end{array}
\right)
\,,
\\ &&
A = 
\left(
\begin{array}{cccc}
-\gamma_{eg} - \gamma_{eq} - \gamma_{ep} & 0 & 0 &0     \\
\gamma_{eg}  &0  & \gamma_{qg} & \gamma_{pg}   \\
\gamma_{eq} & 0 & - \gamma_{qg} -\gamma_{qp} &  0    \\
\gamma_{ep} & 0    & \gamma_{qp}  &   -\gamma_{pg}
\end{array}
\right)
\,, 
\\ &&
\hspace*{-0.5cm}
B_1 = \frac{1}{2}
\left(
\begin{array}{cccccc}
-\Gamma_{eg} &   0  &   -i \Omega_{qg}  &   0 &   0 &   i \Omega_{eq}  \\
0 & -\Gamma_{eg}  &   0  &   i \Omega_{qg}  &   -i\Omega_{eq} &      0 \\
- i \Omega_{qg}     &  0    & -  \Gamma_{eq} & 0 & 0  & 0\\
  0   &   i \Omega_{qg}  & 0 & -  \Gamma_{eq} &0 &0 \\
0 &   -i\Omega_{eq} & 0 & 0& -\Gamma_{qg} &0 \\
i \Omega_{eq} & 0&0 &0 &0 &  -  \Gamma_{qg}
\end{array}
\right)
\,, 
\\ &&
\hspace*{-0.3cm}
B_2 = \frac{1}{2}
\left(
\begin{array}{cccccc}
-\Gamma_{pg} &  0  &   0 &   0 &   i \Omega_{qg} &   0 \\
0  & -\Gamma_{pg} &   0  &   0 &   0 &   - i \Omega_{qg}\\
0    &  0    & -  \Gamma_{ep} & 0&    i \Omega_{eq}  &   0 \\
  0   &0 & 0 & -  \Gamma_{ep}  &   0 &   -  i \Omega_{eq}  \\
 i \Omega_{qg} &  0   &  i \Omega_{eq}  & 0 & -  \Gamma_{pq}  &     0 \\
 0 & -i \Omega_{qg}    &0 & -  i \Omega_{eq}  & 0  &-  \Gamma_{pq} 
\end{array}
\right)
\,, 
\\ &&
\hspace*{-0.3cm}
C_1 = D_1^T = \frac{1}{2}
\left(
\begin{array}{cccccc}
0 &0& i\Omega_{eq} & - i\Omega_{eq} & 0 & 0 \\
 0 & 0 & 0&0&   i\Omega_{qg} &  -i\Omega_{qg} \\
  0&0&- i\Omega_{eq} &  i\Omega_{eq} &  -i\Omega_{qg} &  i\Omega_{qg} \\
0 & 0 & 0& 0 & 0 & 0
\end{array}
\right)
\,, 
\\ &&
C_2 = D_2^T = \frac{1}{2}
\left(
\begin{array}{cccccc}
 0&0 &0 & 0& 0&0\\
- i \Omega_{pg} & i \Omega_{pg} &0 & 0& 0&0\\
 0& 0 & 0& 0& 0&0\\
i \Omega_{pg} &- i \Omega_{pg} &0 &0&0 &0
\end{array}
\right)
\,,
\\ &&
E = F^T = \frac{1}{2}
\left(
\begin{array}{cccccc}
0 & 0 & - i \Omega_{pg} & 0    & 0 & 0   \\
0  & 0  & 0  & i\Omega_{pg}   & 0 & 0 \\
0 & 0 & 0  &  0   & 0 & 0   \\
0 & 0    & 0 & 0  & 0 & 0 \\
0 & 0 & 0 &  0    & 0 &  i \Omega_{pg}   \\
0 &  0  & 0 & 0  & - i \Omega_{pg} & 0
\end{array}
\right)
\,,
\\ &&
\Gamma_{eg} = \gamma_{eg} + \gamma_{eq}  + \gamma_{ep} 
\,, \hspace{0.5cm}
\Gamma_{eq}  = \Gamma_{eg} +\gamma_{qp} + \gamma_{qg}
\,, 
\nonumber \\ &&
\Gamma_{qg} = \gamma_{qp} + \gamma_{qg}
\,, \hspace{0.5cm}
\Gamma_{pg} =  \gamma_{pg}
\,, 
\nonumber \\ &&
\Gamma_{ep}  = \Gamma_{eg} +\gamma_{pg} 
\,, \hspace{0.5cm}
\Gamma_{pq} = \Gamma_{qg}+ \gamma_{pg}
\,.
\end{eqnarray}

\begin{figure*}[htbp]
 \begin{center}
 \epsfxsize=0.8\textwidth
 \centerline{\epsfbox{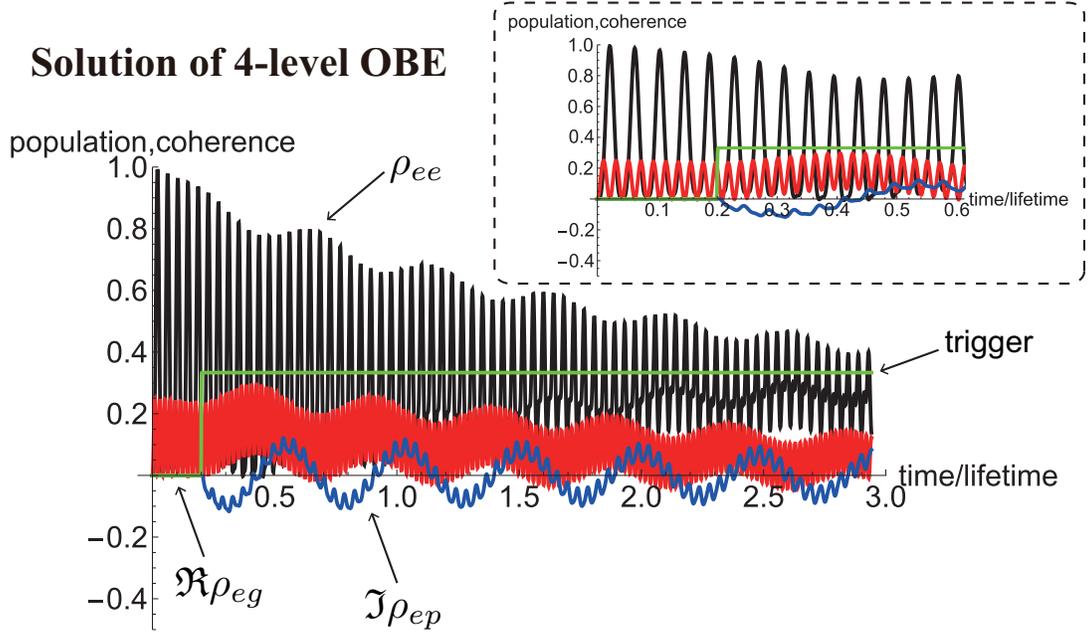}} \hspace*{\fill}\vspace*{1cm}
   \caption{
Solution of four-level optical Bloch equation under CW trigger with a time lag:
$\rho_{ee}$ in black, $\Re \rho_{ eg}$ ($\Im \rho_{ eg} \ll \Re \rho_{ eg}$)
 in red,
and $\Im \rho_{ ep} $  ($\Re \rho_{ ep} \ll \Im \rho_{ ep}$) in  blue.
For reference the trigger profile with time lag is shown in green.
Coherence transfer from $\rho_{eg}$ in red to $\rho_{ep}$ in blue is observed
at the time of trigger irradiation.
Assumed laser intensities are $I_{qg} = 3$mW cm$^{-2}\,, I_{eq} = 9$mW cm$^{-2}\,, I_{pg} = 1\mu$W cm$^{-2}$.
}
   \label {obe 4-level 2}
 \end{center} 
\end{figure*}

We are bound to use for relaxation
 radiative decay rates due to lack of inhomogeneous widths
in actual host crystals.
This might  be a proper choice when we use lowest Stark levels,
but experimental data of relaxation rates of lowest Stark states
in each J-manifold are needed to settle this question.
Time unit used in simulations is the lifetime 1.7 msec of $|e \rangle =$Er$^{3+}\, ^2$H$_{11/2}$.
Calculated Rabi frequencies 
$\Omega_{ab}= d_{ab}E = 2\pi \nu_{ab}$ under CW operation 
of powers $\sim 10 {\rm mW cm}^{-2} = 1.26 \times 10^{-3}$eV$^2$ are of orders,
\begin{eqnarray}
&&
\nu_{pg} = 174 {\rm kHz} \sqrt{\frac{ I_t }{10 {\rm mW cm}^{-2} }}
\,, 
\nonumber \\ &&
\nu_{qg} = 37.4 {\rm kHz} \sqrt{\frac{  I_t}{10 {\rm mW cm}^{-2} }}
\,, 
\nonumber \\ &&
\nu_{eq} = 20.9 {\rm kHz} \sqrt{\frac{ I_t}{10 {\rm mW cm}^{-2} }}
\,,
\end{eqnarray}
using calculated A-coefficients in the text.
Divided by respective radiative decay rates, these $\Omega/\gamma$ are
$(4.0\,, 5.1\,, 4.0) \times 10^4 \sqrt{\frac{  I_t}{10 {\rm mW cm}^{-2} }}$.
Computation of Fig(\ref{obe 4-level 2}) corresponds to use of CW laser power 
of order $\mu$W cm$^{-2}$ or less during  a few msec irradiation.

Solution of large coherence $\rho_{ep}$ generation
is illustrated in Fig(\ref{obe 4-level 2}),
which shows a kind of coherence transfer at the
trigger irradiation, as suggested by heuristic argument using
three-level optical Bloch equation.
Thus, there are ranges of laser parameter  that
give rise to sufficiently large coherence and population.

RENP magnetization emerges with time profile as illustrated in Fig(\ref{renp t-profile}) of the text,
since rates are proportional to $\rho_{ee}^3 |\rho_{ep}|^2$.

\begin{acknowledgments}
This research was partially
 supported by Grant-in-Aid  21H01112(HH), 20H00161(AY), and 
21K03575 (MY)  from the
 Ministry of Education, Culture, Sports, Science, and Technology.

\end{acknowledgments}


\begin{thebibliography}{99}
\bibitem{amplitude formula}
This formula is based on non-relativistic old-fashioned perturbation theory as
adopted in \cite{renp overview}.
\bibitem{e lifetime}
This value arises from the first table of Section \lromn4 where the reference is cited.
\bibitem{even 4f case}
The ground  Stark state of trivalent Ho(4f$^{10}$)
in YLF is a $T=\pm $ doublet of  symmetry $S_4$.
Two lowest Stark singlets of $T=+$ and $T=-$ in upper J-manifolds
may be used for RENP scheme in much the same way as in trivalent Er.
\bibitem{corrected my prd}
We correct a sign mistake of $h_2$ in  pair emission
amplitudes given in the literature \cite{my-07},
eq.(20) and eq.(21).
\end{thebibliography}

\end{document}